\documentclass[aip,pof,amsmath,12pt, amssymb,
preprint%
]{revtex4-1}
\usepackage[utf8]{inputenc}
\usepackage{times,graphics,bm,mathptmx,graphics,
	psfrag,amssymb,amsmath,rotating,color,appendix,mathtools,tikz}
\usepackage{hyperref,natbib}
\usepackage{makecell}
\usepackage{mathtools,siunitx}
\usepackage{graphicx}
\usepackage{hyperref}
\usepackage{titlesec}
\usepackage{amsfonts}
\usepackage{array,booktabs}
\usepackage{multirow}
\usepackage{rotating}

\usepackage{booktabs}
\usepackage{setspace}
\usepackage{soul}
\allowdisplaybreaks
\usepackage[left=2.3cm,right=2.3cm,top=2.3cm,bottom=2.3cm]{geometry}
\hypersetup{
    colorlinks=true, %set true if you want colored links
    linktoc=all,     %set to all if you want both sections and subsections linked
    linkcolor=blue,  %choose some color if you want links to stand out
}

\begin{document}
\title{Penetrative rotating magnetoconvection subject to lateral variations in temperature gradients}
\author{Tirtharaj Barman}
\author{Swarandeep Sahoo}
\email{swarandeep@iitism.ac.in}
\affiliation{Department of Applied Geophysics, Indian Institute of Technology (Indian School of Mines) Dhanbad 826004, India.}

\date{\today}

\begin{abstract}
Convection-driven flows in planetary interiors exhibit rich dynamics owing to multiple spatio-temporally varying forcing conditions and physical constraints. In particular, the churning of liquid metals in the Earth's outer core, responsible for the dynamic geomagnetic field, is subjected to lower mantle thermal heterogeneity. Besides, the plausible existence of a stable stratification layer below the mantle influences the columnar convection. These additional symmetry-breaking constraints, motivated from geophysical scenario of the Earth's thermal core--mantle interaction, modulate the otherwise periodic and axially invariant convection flow patterns. Thus, the present study focuses on qualitative characterization and parametric quantification of rotating penetrative convection in the presence of magnetic induction effects with an aim to understand the role of the lower mantle on core convection thermally. Using complementing computational and theoretical calculations, the present study estimates the depth of penetration in bounded and unbounded fluid domains. Apart from qualitative differences in convective flow patterns from the reference homogeneous configurations, the additional constraints spatially modulate the extent of penetration into stable regions. Confinement effects, adding to the damping of penetrative convection, arising out of boundary constraints are quantified for bounded geometry. Appropriate normalizations, implemented to eliminate such effects, result in amended penetration depth estimates that align with the qualitative characteristics obtained for unbounded domains. Exact closed-form expressions for the depth of penetration are obtained, providing insights into the role of individual contributions of multiple physical constraints. Implications are speculated for realistic, yet unreachable, regimes of geophysical conditions of planetary cores.
 
\end{abstract}

\maketitle

\section{Introduction}
Various geophysical investigations such as seismology \cite{kaneshima2018array, lay1990stably} and geomagnetism \cite{buffett2014geomagnetic, buffett2010stratification, gubbins2007geomagnetic} have inferred a possible existence of chemical\cite{helffrich2010outer} and thermal\cite{greenwood2021evolution} stable stratification at the top of the Earth's outer core. Additionally, from mineral physics studies, the first principal calculations on band structure\cite{gomi2018impurity} as well as thermal and electrical conductivity\cite{pozzo2012thermal} of $Fe$ alloy in core condition favor the existence of thermally stable stratification just below the Earth's core-mantle boundary ($CMB$). Magnetic-Archimedes-Coriolis ($MAC$) waves occurring in the presence of such stable stratification are characterized by oscillation periods corroborating the fluctuation in geomagnetic secular variations ($SV$)\cite{buffett2014geomagnetic}. Based on geophysical constraints, several theoretical, numerical, and experimental approaches have been proposed to study the mechanism behind forming a stable layer below $CMB$. Thermally stable stratification may arise if the heat flow at the $CMB$ falls below the adiabatic heat flow at the core\cite{gubbins1982stable, lister1998stratification}. On the other hand, a chemically stable layer may be developed due to baro-diffusion of light elements ($Si, O, Mg$) across the fluid outer core \cite{gubbins2013stratified} leading to the accumulation of distinct light elements at the $CMB$ \cite{bouffard2019chemical}. Extensive studies concerning the effects of partial stable stratification on convective\cite{dietrich2018penetrative} and dynamo processes\cite{mukherjee2023thermal} related to the Earth and other planetary bodies\cite{moore2022dynamo,gastine2021stable} are noteworthy. Fundamental insight into the impact of partial stable stratification on the onset of rotating thermal convection \cite{garai2022convective}, rotating magnetoconvection\cite{sreenivasan2024oscillatory, xu2024penetrative} and thermo-fluidic characteristics in strongly driven turbulent convection\cite{barman2024role} have been the focus of numerous investigation using plane layer models. 

Based on several factors convective flows can penetrate from an unstable layer to an adjacent stable layer, known as penetrative convection\cite{veronis1963penetrative}. In presence of rapid rotation \cite{garai2022convective, barman2024role}, magnetic field \cite{xu2024penetrative, sreenivasan2024oscillatory}, strong buoyancy forcing\cite{barman2024role, toppaladoddi2018penetrative} the penetrative convection models exhibits rich dynamics\cite{dietrich2018penetrative, mukherjee2023thermal}. Under rapid rotation regime, the penetration extent into stable layer of convective flows may enhance \cite{garai2022convective, barman2024role}, however, temperature fields get confined into the convectively unstable regime \cite{garai2022convective, barman2024role}. The incorporation of a background magnetic field into the fluid system imparts additional stability and results in substantial modification in convective flows based on the choice of orientation of the magnetic field and its strength.\cite{xu2024penetrative, sreenivasan2024oscillatory}. Further incorporation of laterally varying thermal structure may induce additional confinement of flow to higher heat flow regimes than the mean flow \cite{garai2022convective}, that may be modified in presence of rapid rotation for plane layer model \cite{garai2022convective} and combined effect of magnetic field and rapid rotation for f-plane model\cite{xu2024penetrative} and spherical shell model of magnetoconvection \cite{olson1996magnetoconvection}. 

Geometrical modeling of the Earth's outer core assumes a spherical shell containing an inner sphere and an outer shell representing the solid inner core and the $CMB$, respectively, encompassing the fluid outer core (Fig. \ref{f_schematic}($a$)). Denoting the radii of the outer and inner boundaries by $r_o$ and $r_i$, respectively, the radius $r = r_s$ marks the interface separating stable stratification above ($r_s < r < r_o$)  and unstable stratification below ($r_i < r < r_s$). Numerical modeling of convection and consequent dynamo action in such spherical shell models depend on the chosen buoyancy profiles\cite{davies2011buoyancy} mimicking super-adiabatic condition in the outer core. Prominently, outer core convection is driven by thermal gradients arising out of inner core freezing (differential heating profile\cite{sahoo2017effect}) 
and secular cooling (internal heating profile) processes\cite{dormy2004onset, sahoo2017effect}. The above fundamental buoyancy profiles may be augmented to accommodate partial thermal or chemical stratification\cite{bouffard2019chemical} by the inclusion of additional physical constraints such as radial heat source variations\cite{mukherjee2023thermal} modified by lateral heterogeneity in boundary heat flux \cite{christensen2018geodynamo,garai2022convective,barman2024role}, interior heat sources and spatially non-uniform magnetic fields\cite{sreenivasan2024oscillatory}.

Under rapid rotation, convection in the Earth's core is separated into two regions by the imaginary tangent cylinder ($TC$), whose axis is tangential to the inner core at the equator, intersecting the $CMB$ at $70^{\circ} N-S$ latitudes\cite{olson2015core} (Fig. \ref{f_schematic}($a$)). The convective dynamics inside and outside $TC$ are significantly different owing to the alignment of various influencing forces\cite{jones2015thermal,aurnou2003experiments}. The present study focuses on the dynamics inside the TC where the buoyancy flux has substantial components along the axial direction. Consequently, the fluid motion becomes strongly dependent on the axial direction requiring ageostrophic motions to transport heat and light element from the inner core surface to $CMB$\cite{jones2015thermal}. Considering the dynamic similarity due to prevailing alignment of forcing factors and neglecting curvature effects for simplicity, a plane layer model is employed\cite{jones2000convection} in the present study to avoid large computational costs associated with full sphere models. Besides that, several previous studies have also considered unbounded fluid domains\cite{finlay2008course, finlay2005hydromagnetic} to explore slow and fast magneto-hydrodynamic wave characteristics and their implication to dynamo action\cite{luo2022waves} and convection inside $TC$ \cite{sreenivasan2021evolution,varma2022role,majumder2024self, majumder2023role} and geomagnetic secular variations\cite{finlay2005hydromagnetic} motivating its implementation in this study.

The convective flows originating from thermally unstable regions can penetrate the stable layer\cite{veronis1963penetrative}. Based on the strength of stratification, buoyancy forcing, rotation rate, strength and orientation of imposed magnetic field, and fluid properties the penetration depth may vary substantially. The penetration depths have been estimated for spherical shell models \cite{takehiro2001penetration, takehiro2015penetration, takehiro2018penetration, vidal2015quasi} and f-plane models\cite{xu2024penetrative} using constant\cite{takehiro2001penetration} and linearly varying stratification profiles \cite{vidal2015quasi}. In unbounded fluid domain, theoretical expressions for penetration depth have been derived for hydromagnetic waves\cite{takehiro2015penetration,takehiro2018penetration}. 
In numerical simulations, the depth of penetration of convective flows has been estimated either by measuring the $e-$ folding drop distance of axial velocity\cite{takehiro2001penetration} or by estimating distance between the $10\%$ drop of convective flux\cite{gastine2020dynamo,xu2024penetrative}. However, the penetration depth estimates deviated from the theoretically obtained values for configurations where the penetration depth became larger than thickness of stable layer. By using transmission coefficients this ambiguity was relaxed to a certain extent\cite{vidal2015quasi}. Moreover, penetrative convection models have also attempted to estimate the thickness of stable stratification at the top of the Earth's core \cite{gastine2020dynamo}.

The present study aims to identify the characteristics of penetrative flows under multiple geophysical constraints relevant to the Earth's outer core by amalgamating both numerical and analytical techniques. The relevant configurations along with geometrical and mathematical modeling are presented in section \ref{sec2}. The detailed results for the bounded and unbounded domains are given in section \ref{sec3}. A comparative discussion and relevant conclusions are provided in section \ref{sec4}. Step-wise details of analytical derivations can be found in the appendices \ref{sec5}.

\section{Formulation}\label{sec2}
\subsection{Problem setup}\label{sec2.1}
\begin{figure}[!]
    \centering
    \includegraphics[clip, trim=0cm 15cm 0cm 0cm, width=1\textwidth]{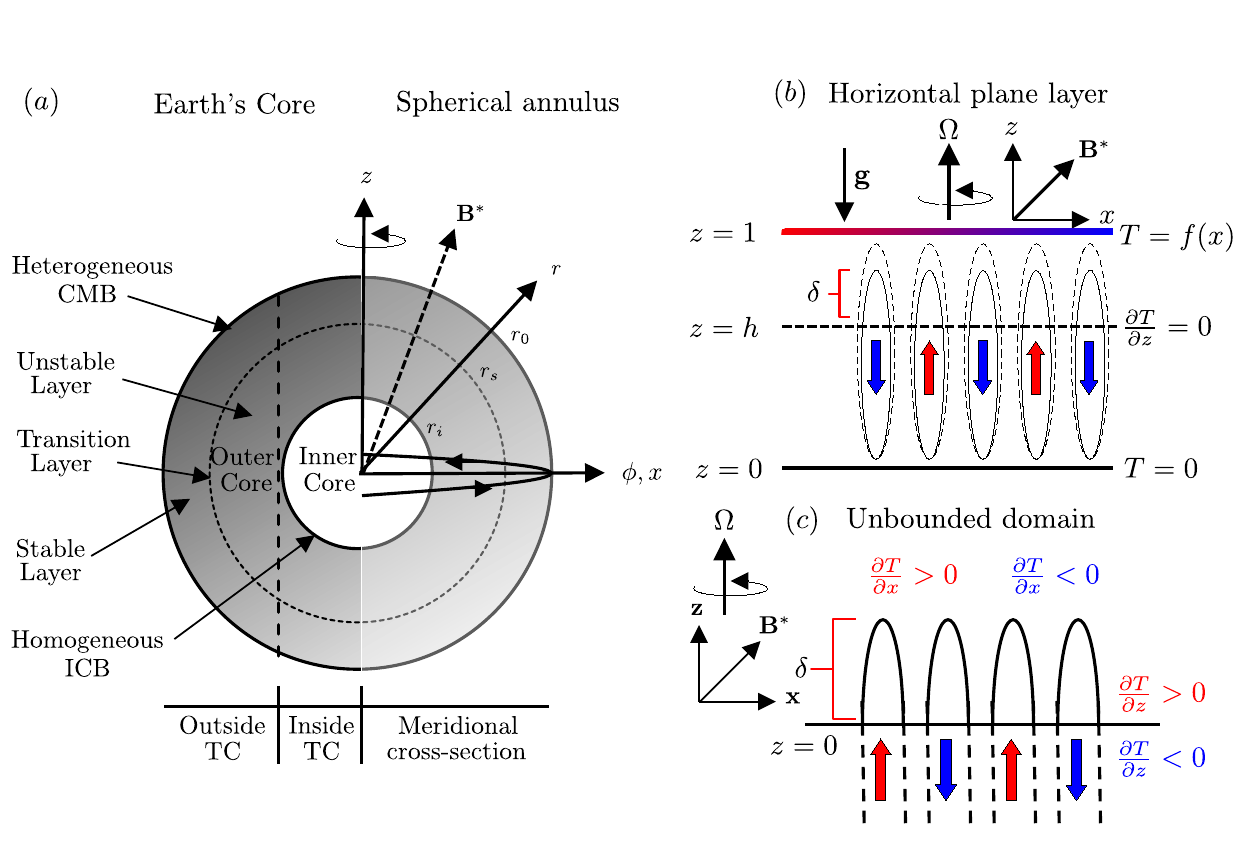}
    \caption{Schematic diagrams of ($a$) spherical shell, (left) internal structure of the Earth and (right) meridional cross-section in the spherical coordinate system; ($b$) confined rotating plane layer; ($c$) unbounded fluid.}
    \label{f_schematic}
\end{figure}

The study investigates two geometrically different fluid dynamical models, appropriately complementing the physical constraints and theoretical insights. The first is a horizontally infinite plane layer with bounding surfaces along the vertical ($z$) direction only separated by a gap-width $D$. [Fig.\ref{f_schematic}(b)]. The second configuration is of an unbounded fluid domain, extending to infinity in all the three ($x,y,z$) directions [Fig.\ref{f_schematic}(c)]. The Cartesian coordinate system with the vertical axis ($z$) and lateral axes ($x,y$) has been used in the current study with the lengths scaled by $D$. Background rotation is imposed with a rotation rate $\bm{\Omega} = \Omega_0 \hat{z}$ over the complete domain. Thermal forcing enters the system through temperature gradients ($\frac {\partial T}{\partial z}$) and gravity ($\mathbf{g} = -g \hat{z}$). In the plane layer model, the hotter bottom ($z = 0$) and colder top ($z = 1$ ) boundaries are fixed in temperature, resulting in the axial temperature gradients while a background thermal gradient is imposed for the unbounded domain configuration. Additionally, for magnetoconvection, a uniform background magnetic field is imposed either along the horizontal ($B^* = \hat{x}$) or vertical ($B^* = \hat{z}$) direction. The above configuration represents the homogeneous reference case with axial and horizontal uniformity of both components of the temperature gradients ($\frac {\partial T}{\partial z}$ and $\frac {\partial T}{\partial x}$).

In order to incorporate additional physical constraints relevant to the Earth's outer core dynamics, the symmetrical reference configurations described above are modified appropriately. Firstly, a stably stratified layer ($\ \frac {\partial T}{\partial z} > 0$), believed to exist at the outer regions of the Earth's core ($r_s < r < r_o$), is introduced above a thermally unstable ($\ \frac {\partial T}{\partial z} < 0$) fluid region. In the plane layer configuration, the stable layer occurs for $h < z < 1$, with {$z-h$ being the interface between the thermally stable and unstable regions [Fig. \ref{f_schematic}(b)]. For the unbounded domain configuration, the corresponding stable-unstable interface occurs at the $x-y$ plane denoted by $z = 0$ [Fig.\ref{f_schematic}(c)]. In the context of the Earth's core, the stable and unstable regions correspond to sub-adiabatic and super-adiabatic regions \cite{lister1998stratification} respectively. To aid in understanding, the damping effects of stable layers \cite{garai2022convective, barman2024role} is schematically represented with overlying solid (damped) and broken (undamped) circulatory fluid motions. 

Besides the stable stratification, thermal core-mantle interaction due to lateral variations in lower mantle heat flow is a significant physical constraint that is implemented in this study. To this end, the imposed temperature at the top boundary is spatially modulated as $T = f(x)$ at $z = 1$ (figure \ref{f_schematic}(b)). Correspondingly, lateral variations in the background temperature are imposed for the unbounded domain configuration. The additional physical constraints of partial stable layer and lateral temperature variations break the axial\cite{barman2024role} and lateral symmetry\cite{garai2022convective}, respectively, of the corresponding reference configurations. In other words, the axial and horizontal uniformity of the reference case is lost due to partial stable stratification and upper boundary temperature variations.

The most appropriate configuration relevant to the Earth's outer conditions includes the combined effects of partial stable stratification with lateral modulations in buoyancy forcing\cite{christensen2018geodynamo}. Such a configuration leads to regional thermal stratification\cite{garai2022convective} which can plausibly influence the penetrative convection instabilities resulting in spatially varying penetration depth. The present study aims to obtain the quantitative estimates and qualitative aspects of penetration depth for dynamically relevant conditions as appropriate for the Earth's core scenario. However, due to the enormous computational resources necessary to perform a systematic study in a spherical shell, the present study investigates a geometrically simplified, yet physically relevant plane layer model \cite{garai2022convective, barman2024role}. Furthermore, the symmetry-breaking constraints restrict the plane layer study to numerical simulations. Thus, the investigation is complemented with the configuration of an unbounded fluid domain which is amenable to a theoretical approach and provides deeper physical insights resulting from closed-form solutions, not obtainable from discrete computational modeling.

\subsection{Governing equations}\label{sec2.2}
To study the above thermo-fluidic system, an incompressible fluid is considered with density perturbations prevailing only for the buoyancy forcing. In this Boussinesq limit, the density changes can be represented by  $\rho = - \rho_0 \alpha T$, $\rho_0$ being the background density. Thus, the dimensional form of the governing equations is given by \cite{kundu2015fluid},
\begin{equation} \label{eq_incom_magnetic}
\mathbf{\nabla} \cdot \mathbf{u} = 0,
\end{equation}
\begin{equation} \label{eq_monopole_magnetic}
\mathbf{\nabla} \cdot \mathbf{B} = 0,
\end{equation}
\begin{equation} \label{eq_NS_magnetic}
\frac{\partial \mathbf{u}}{\partial t} + (\mathbf{u} \cdot \nabla)\mathbf{u} + 2 \Omega_0\hat{\textbf{z}} \times \mathbf{u} = -\frac{\nabla P} {\rho_0} + \alpha T g \hat{\textbf{z}} + \frac{1}{\rho_0 \mu_0} (\nabla \times \textbf{\textit{B}}) \times \textbf{\textit{B}} + \nu \nabla^2 \mathbf{u},
\end{equation}
\begin{equation} \label{eq_temp_magnetic}
\frac{\partial T}{\partial t} + (\mathbf{u} \cdot \nabla) T  = \kappa \nabla^2 T + Q
\end{equation}
\begin{equation} \label{eq_induction_magnetic}
\frac{\partial \textbf{\textit{B}}}{\partial t}  = \textit{$\nabla$} \times (\textbf{\textit{u}} \times \textbf{\textit{B}}) + \eta \nabla^2 \textbf{\textit{B}}
\end{equation}
where, $\textbf{u}$ is the  velocity vector and $\textbf{B}$ is the magnetic field vector. $Q$ is the volumetric internal heating rate representing the secular cooling of the Earth's outer core fluid. 

The modified pressure gradient $\nabla P$ = $\nabla P_{H} - \frac{1}{2} \nabla |\bm{\Omega} \times \mathbf{r}|^2 +  \frac{1}{2} \nabla (\mathbf{B} \cdot \mathbf{B})$ includes centrifugal acceleration and magnetic pressure gradient besides the gradient of hydrodynamic pressure ($P_H$). The coefficients of thermal expansion, viscous diffusivity, thermal diffusivity, and magnetic diffusivity are denoted by  $\alpha$, $\nu$, $\kappa$, and $\eta$, respectively. All the above fluid properties are assumed to be constant. 

Upon Reynolds decomposition, all the components (if any) of the field variables among [$\mathbf{u}$, $\mathbf{B}$, $T$ and $P$] denoted by $\mathbf{\psi}$ can be represented by their steady basic states (indicated by superscript $*$) and time-dependent perturbation state (indicated by superscript $\prime$) as,
\begin{equation} \label{eq_bi_global_total}
\mathbf{\psi} = \mathbf{\psi}^*(x,y,z) + \mathbf{\psi}^{\prime}(x,y,z,t)
\end{equation} 

In the absence of any background magnetic field, a purely thermal convection model is described by the reduced system of equations given by equation (\ref{eq_incom_magnetic}), equation (\ref{eq_NS_magnetic}) without the Lorentz term and equation (\ref{eq_temp_magnetic}) only. These equations are scaled using $D$ as the length scale, $\frac{D^2}{\kappa}$ (thermal diffusion time) as the time scale, $\Delta T = T_1 - T_0$ 
%the applied temperature difference between two layers,  
as the temperature scale and \textbf{$\rho \frac{\eta^2}{D^2}$} is dynamic pressure as the scale for pressure. 

Assuming a static steady basic state ($u^* = 0$ and $\frac{\partial}{\partial t} = 0$), the perturbation state equations for penetrative thermal convection are obtained as,
\begin{equation} \label{eq_incom_dimensionless_perturb}
\mathbf{\nabla} \cdot \mathbf{u^{\prime}} = 0,
\end{equation}
\begin{equation} \label{eq_NS_dimensionless_perturb}
\frac{\partial \mathbf{u^{\prime}}}{\partial t} + (\mathbf{u^{\prime}} \cdot \nabla)\mathbf{u^{\prime}} + \frac{Pr}{E} (\mathbf{\hat{z}} \times \mathbf{u^{\prime}}) = - \nabla P^{\prime} + RaPr T^{\prime}  \hat{\textbf{z}} + {\color{red}Pr}\nabla^2 \mathbf{u^{\prime}},
\end{equation}
\begin{equation} \label{eq_temp_dimensionless_perturb}
\frac{\partial T^{\prime}}{\partial t} + (\mathbf{u^{\prime}} \cdot \nabla) T^{\prime} + u_x^{\prime} \Gamma_x^* + u_z^{\prime} \Gamma_z^* = \nabla^2 T^{\prime}
\end{equation}
where the subscripts ($x$ or $z$) denote the corresponding vector component. The control parameters are dimensionless numbers, namely, the Rayleigh number ($Ra$), thermal Prandtl number ($Pr$) and Ekman number ($E$) defined as,
\begin{equation} \label{eq_parameters_nonmag}
Ra = \frac{g \alpha \Delta T D^3}{\nu \kappa}, \hspace{15pt} Pr = \frac{\nu}{\kappa}, \hspace{15pt} E = \frac{\nu}{2 \Omega_0 D^2}
\end{equation}
The Rayleigh number ($Ra$) is the ratio of thermal buoyancy forces to viscous drag, which determines the strength of convection. Thermal Prandtl number ($Pr$) measures the relative importance between viscous diffusion and thermal diffusion. Ekman number ($E$) indicates the ratio of viscous to Coriolis forces. A lower $E$ regime indicates stronger rotational effects.

With augmentation to magneto-convection configuration, the working fluid is further assumed to have finite electrical conductivity. Incorporation of magnetic effects necessitates the non-dimensionalization of the full governing equations (\ref{eq_incom_magnetic}) - (\ref{eq_induction_magnetic}) by using the magnetic diffusion time ($\frac{D^2}{\eta}$) as the time scale, $\frac{\eta}{D}$ as velocity scale and the imposed field strength ($B_0$) as the magnetic field scale in addition to other scales used for thermal convection system. Hence, the perturbation state equations for penetrative magneto-convection are obtained as,
\begin{equation} \label{eq_incom_magnetic_dimensionless_perturb}
\nabla \cdot \mathbf{u^{\prime}} = 0 , \hspace{10pt} \nabla \cdot \mathbf{B^{\prime}} = 0,
\end{equation}
\begin{equation}\label{eq_NS_magnetic_dimensionless_perturb}
\begin{split}
\frac{\partial \mathbf{u^{\prime}}}{\partial t} + (\mathbf{u^{\prime}} \cdot \nabla) \mathbf{u^{\prime}} + \frac{Pm}{E} (\mathbf{\hat{z}} \times \mathbf{u^{\prime}}) = -\nabla \bar{P^{\prime}} +   \frac{\Lambda Pm}{E} (\mathbf{B^*} \cdot \nabla) \mathbf{B^{\prime}}  \\   + \frac{\Lambda Pm}{E} (\mathbf{B^{\prime}} \cdot \nabla) \mathbf{B^{\prime}} + q Ra Pm T^{\prime} \hat{\mathbf{z}} + Pm \nabla^2 \mathbf{u^{\prime}},
\end{split}
\end{equation}
\begin{equation} \label{eq_temp_magnetic_dimensionless_perturb}
\frac{\partial T^{\prime}}{\partial t} + (\mathbf{u^{\prime}} \cdot \nabla) T^{\prime} + u_x^{\prime} \Gamma_x^* + u_z^{\prime} \Gamma_z^* = q \nabla^2 T^{\prime},
\end{equation}
\begin{equation} \label{eq_induction_magnetic_dimensionless_perturb}
\frac{\partial \mathbf{B^{\prime}}}{\partial t}  = (\mathbf{B^{\prime}} \cdot \nabla) \mathbf{u^{\prime}} + (\mathbf{B^*} \cdot \nabla) \mathbf{u^{\prime}} + \nabla^2 \mathbf{B^{\prime}}
\end{equation}
Here, ${P}^{\prime} = P^{\prime} - \frac{Pr}{E} |\mathbf{\hat{z}} \times \mathbf{r}| +  \frac{\Lambda Pm}{E} \nabla (\mathbf{B}^{\prime} \cdot \mathbf{B}^{\prime})$ is the modified pressure gradient perturbation. 
%includes centripetal acceleration and magnetic pressure. Where, new 
The additional dimensionless control parameters are magnetic Prandtl number ($Pm$), Elssaser number ($\Lambda$), and Roberts number ($q$), defined as,
\begin{equation} \label{eq_parameters_mag}
Pm = \frac{\nu}{\eta}, \hspace{15pt} \Lambda = \frac{B_0^2}{2\Omega \rho \mu \eta }, \hspace{15pt} q = \frac{Pm}{Pr}
\end{equation}
The magnetic Prandtl number ($Pm$) measures the relative significance of viscous diffusion over magnetic diffusion. The Elssaser number ($\Lambda$) is the ratio of Lorentz forces to Coriolis forces. The relative importance of thermal diffusion to magnetic diffusion is measured by Roberts number ($q$). $\Gamma_z^* = \frac{\partial T^*}{\partial z}$ and $\Gamma_x^* = \frac{\partial T^*}{\partial x}$ are the dimensionless basic steady state axial ($z$) and lateral ($x$) temperature gradients, respectively. \\
Although far from Earth-like conditions, the parameter regimes investigated retain the dynamical similarity in terms of the hierarchy of forcing mechanisms \cite{guzman2021force}. Thus, the dominance of rotational effects ($E << 1$) followed by Lorentz (if applicable), buoyancy forcing is maintained. The inertial and viscous forces are kept to a minimum. The values of $E$ range from $10^{-4} - 10^{-3}$, $Pr$ range from $0.01, 0.1, 1$, and for magneto-convection, $q = 1$. The strength of the imposed magnetic field is chosen as $\Lambda = 0.01$ (weak field) and $\Lambda =5$ (strong field) regimes.

\subsection{Methods} \label{sec2.3}

In the present study, both computational modeling and theoretical analysis have been implemented to obtain phenomenologically relevant and physically insightful results for penetrative rotating convection. Direct numerical simulations (DNS) have been performed for the plane layer configuration. This involves discretization of the relevant perturbation equations for thermal convection (Eqs. $(\ref{eq_incom_dimensionless_perturb}) - (\ref{eq_temp_dimensionless_perturb})$) and magnetoconvection (Eqs.$(\ref{eq_incom_magnetic_dimensionless_perturb}) - (\ref{eq_induction_magnetic_dimensionless_perturb})$). No-slip and fixed temperature boundary conditions\cite{garai2022convective} $(u^{\prime}_x, u^{\prime}_z, T^{\prime}) = (0, 0, 0)$ are applied for velocity and temperature perturbations at $z = 0, 1$. For magnetoconvection, pseudo-vacuum\cite{jackson2014spherical} condition $(B_x, \frac{\partial B_z}{\partial z}) = (0,0)$\cite{jones2000convection} is implemented for magnetic field perturbation at $z = 0, 1$. The basic states are obtained using solutions to the steady state equations with relevant boundary conditions\cite{garai2022convective,barman2024role}. \\
The governing equations are numerically solved using spectral methods with the boundary conditions implemented using the spectral Tau method using the Dedalus library \cite{burns2020dedalus}. The spatial discretizations along the vertical and horizontal directions are performed using Chebyshev and Fourier series expansions, respectively. Spectral resolutions up to a maximum of $256 \times 256$ spectral coefficients are utilized for ensuring spectral accuracy. A $2^{nd}$ order Runge-Kutta method has been implemented with a maximum time step of $10^{-4}$ for temporal evolution with verified convergence and stability. \\
The spatial non-uniformity in the basic thermal state (Eqs.\ref{eq_bi_global_total}) necessitates the use of global normal mode analysis\cite{taira2020modal} for the unbounded fluid configuration. The global analysis assumes a general functional form for the perturbations in the spatial directions ($x,y,z$) and wave-like form in time $t$ given by, 
\begin{equation} \label{eq_bi_global_perturbation}
 {u_z}^{\prime}(x,y,z,t) = {u_z}^{\prime}(x,y,z) e^{i\zeta t}. 
\end{equation}
Here, $\zeta$ is the temporal frequency of the perturbation. The real and imaginary parts of the complex frequency $\zeta (= \zeta_r + I \zeta_i$) represent the angular frequency and growth ($\zeta_i > 0$)/damping ($\zeta_i < 0$) rate of the perturbation amplitude, respectively. However, in the present study, only neutral stationary modes are considered denoted by $\zeta_i = 0$, hence $\zeta = \zeta_r$.  
Subsequent analytical derivations of the penetration depth involve obtaining dispersion relations from relevant perturbation equations. 

\section{Results}\label{sec3}
\subsection{Penetrative convection in confined fluid layer}\label{sec3.1}
The convective flows driven by buoyancy forcing, when subjected to stable stratification, are inherently restricted spatially \cite{garai2022convective, barman2024role}. Furthermore, the confinement effect of impenetrable bounding surfaces also inhibits the axial velocity ($u_z^{\prime}$), a major component of rotationally dominant convective flows. A significant aspect of such configurations is the steady state buoyancy profile ($T^{\star}(z)$), which not only dictates the threshold of the convective onset but also plays a role in the spatio-temporal structure of the convective instabilities. The penetrative effects on fluid flows, in vertically bounded plane layers driven by rotating thermal convection with and without the presence of background magnetic fields, due to both regional stratification and geometrical confinement are investigated below.

\subsubsection{Buoyancy profiles}
The fluid flow subjected to convection is influenced by the imposition of super-adiabatic temperature gradient along the axial direction. Using a combination of heat source/sink in the interior of the fluid layer a non-linear temperature profile is developed to implement regional stratification, following previous fundamental studies \cite{matthews1988model}. On top of that, a sinusoidally varying temperature boundary condition is imposed on the top boundary.

From equation (\ref{eq_temp_magnetic}), under static ($\mathbf{u}^*$) and steady state ($\frac{\partial}{\partial t} = 0$) assumption, the basic state temperature equation is obtained as,
\begin{equation} \label{eq_basic_temperature_thermal}
\nabla^2 T^* = - Q.
\end{equation}
Configurations with  heterogeneous temperature ($T^* = sinx$) condition at the top boundary and constrained by $\frac{\partial T^*}{\partial z} = 0$ at $z = h$, the basic state temperature profile is obtained as,
\begin{equation} \label{eq_basic_temperature_profile_thermal}
T^*(x,z) = -\frac{Q z^2}{2} (1-sinx) + \frac{Q-2}{2}z(1-sinx) + 1,
\end{equation}
where, $Q = \frac{2}{1-2h}$. For homogeneous  isothermal ($T^* = 1$) top boundary, the  basic state temperature profile simplifies to
\begin{equation} \label{eq_basic_temperature_profile_thermal_differential}
T^*(x,z) = 1- z
\end{equation}

The magnetoconvection system with the incorporation of a background magnetic field and electrically conducting fluid, the corresponding basic state temperature  equation (\ref{eq_basic_temperature_thermal}) is modified to,
\begin{equation} \label{eq_basic_temperature_magnetic}
\nabla^2 T^* = - \frac{Q}{q}
\end{equation}
and the corresponding basic state temperature profile with $T^* = sinx$ at $z = 1$ is obtained as,
\begin{equation} \label{eq_basic_temperature_profile_magnetic}
T^*(x,z) = -\frac{Q z^2}{2q} (1-sinx) + \frac{Q-2q}{2q}z(1-sinx) + 1.
\end{equation} 
Due to the presence of the magnetic effects, the thermal diffusion becomes dependent on the Roberts number ($q$) 
such that the source/sink term becomes $Q = \frac{2q}{1-2h}$. As $q = 1$ for all cases, the source/sink distribution and basic state temperature profiles remain the same for the magnetic and non-magnetic cases. 
In the above configurations, the basic state axial ($z$) and lateral ($x$) temperature gradients are obtained as,
\begin{equation} \label{eq_basic_temperature_gradient_Z}
\frac{\partial T^*}{\partial z} = \frac{2(h-z)}{1-2h}(1-sinx)
\end{equation}
and
\begin{equation} \label{eq_basic_temperature_gradient_X}
\frac{\partial T^*}{\partial x} = \frac{z^2 - 2hz}{1-2h}cosx
\end{equation}
respectively. The temperature gradients vary laterally due to the heterogeneous thermal boundary condition implemented at the top boundary. 

The penetrative depth of convective instabilities depends on the strength of the stable stratification\cite{takehiro2001penetration}. The strength of stable stratification is quantifiable based on the axial temperature gradients. This is performed using the Brunt-Väisälä frequency defined as,
\begin{equation} \label{eq_Burnt_vaisala_z}
N = \sqrt{\alpha g \frac{\partial T}{\partial {z}}},
\end{equation}
For thermally stable stratified layers, $N>0$.
%$[{N_x}^2, {N_z}^2] > 0$ and vice-versa. 
The dimensionless form of the  Brunt-Väisälä frequency is obtained by appropriate scaling as,
\begin{equation} \label{eq_Burnt_vaisala_thermal_z}
%N_{x_i}^* = \sqrt{Ra Pr \Gamma_{x_i}^*}.
N_{z}^* = \sqrt{Ra Pr \Gamma_{z}^*}.
\end{equation}
The above expression is also applicable to magnetoconvection cases. As the temperature gradients are axially non-constant over the stable layer extent, for practical purposes, the effective temperature gradients ($\Gamma^{\star}$) are used to estimate the overall penetration depths. The effective temperature gradient is estimated using integrally averaged values across the stable layer ($z > h$) given by,
\begin{equation}
\label{eq_gammaZ-X_avg}
\Gamma_z^* (x,h) = \frac{1}{1-h} \int_{h}^{1} \frac{2(h-z)}{1-2h}(1-sinx) \,dz ,  \hspace{15pt}  \Gamma_x^* (x,h) = \frac{1}{1-h} \int_{h}^{1} \frac{cosx(z^2 - 2hz)}{1-2h} \,dz 
\end{equation}
Thus for the stable layer region ($z>h$), the effective axial and horizontal temperature gradients are expressed as,
\begin{equation} \label{eq_gammaZ-X}
\Gamma_z^* (x,h) = \frac{(1-h)(1-sinx)}{2h - 1},  \hspace{15pt}  \Gamma_x^* (x,h) = \frac{cosx(1-3h+2h^3)}{3(1-3h+2h^2)} 
\end{equation}

\begin{figure}[htbp]
    \centering
    \includegraphics[clip, trim = 0cm 22cm 0cm 0cm, width=1\textwidth]{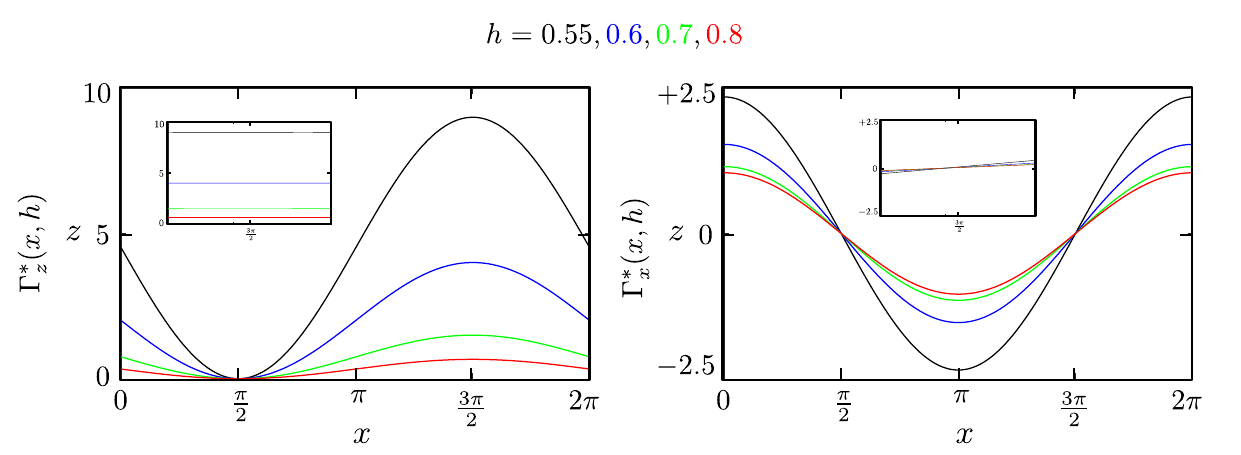}
    \caption{Lateral ($x$) profiles of axially averaged (a) vertical temperature gradient ($\Gamma_z^*$), and (b) horizontal temperature gradient ($\Gamma_x^*$). Black, Blue, Green, and Red solid lines indicate the stratification height of $h = 0.55, 0.6, 0.7, 0.8$, respectively. The inset represents horizontally constant gradients near $x = \frac{3 \pi}{2}$.}
    \label{f_Gamma}
\end{figure}

The horizontal profiles for the effective axial and lateral temperature gradients for various values of $h$ (Fig.\ref{f_Gamma}) characterize the strength of the stable stratification modified by imposed buoyancy heterogeneity. Axial gradients (Fig.\ref{f_Gamma}a) attain maximum positive values at $x = 3\pi/2$ corresponding to the most stable region while at $x = \pi/2$, neutral conducting regime persists irrespective of $h$. The lateral temperature gradients (Fig.\ref{f_Gamma}b) are ineffective at these locations. Note that the regions of negative $\Gamma^{\star}_z $ between $\pi < x < 3\pi/2$ and positive gradient, elsewhere, may lead to transverse ($u_y$) steady baroclinic shear flows. The magnitude of the gradients decrease rapidly with enhancement in $h$ asymptotically converging to negligible values as $h \rightarrow \infty$.

\subsubsection{Penetrative thermal convection}\label{sec3.2}
The onset of convection in the plane layer occurs at a threshold buoyancy forcing also termed as the critical Rayleigh number, $Ra_c$. 
The critical parameters such as onset wave number and frequency are modified substantially, from the corresponding reference case values \cite{chandrasekhar2013hydrodynamic}, in the presence of stable stratification and heterogeneous thermal boundary conditions \cite{garai2022convective}. 

As convective flows are generated in the thermally unstable region the presence of a stable stratification layer inhibit their axial penetration. The penetration extent varies based on the strength of the stable stratification and the confinement effects of impenetrable top boundary. Quantitative estimation and qualitative characterization of the penetration depth are performed over an extensive parameter regime.  
 
\begin{figure}[htbp!]
\centering
\includegraphics[clip, trim = 0cm 20cm 0cm 0cm, width=1\textwidth]{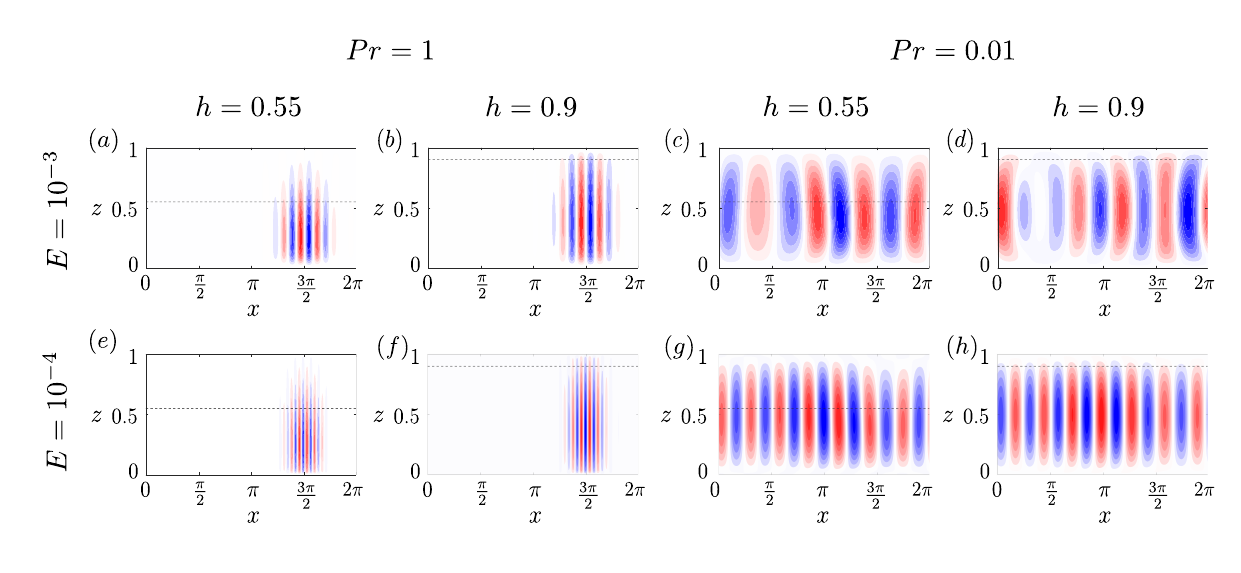}
\caption{(a)-(h) are contour maps of axial velocity ($u_z^{\prime}$) at onset for penetrative thermal convection. (a)-(d) indicate moderate rotation case, $E = 10^{-3}$ and (e)-(h) indicate rapid rotation, $E = 10^{-4}$. (a,e) for interface height at $h = 0.55$, (b,f) for interface height at $h = 0.9$ at $Pr = 1$ and (c,g) for interface height at $h = 0.55$, (d,h) for interface height at $h = 0.9$ at $Pr = 0.01$.}
\label{f_nonmag_contour}
\end{figure}

Axial velocity ($u_z^{\prime}$) with different rotation rates and diffusivity ratios are considered for strong to weak stratification (Figs.\ref{f_nonmag_contour}). Imposition of lateral in-homogeneity and stably stratified layer near the top boundary results in axial and lateral confinement of convective flows. In general fluid flow remains confined in $\pi < x < 2\pi$ due to the influence of temperature variations imposed at the top boundary. Due to this imposition, the maximum axial temperature gradient occurs at $x = \frac{3 \pi}{2}$  leading to clustering of columnar convective instabilities around the maximal thermal gradients. On the other hand, the horizontal location $x = \frac{\pi}{2}$ corresponds to an almost negligible axial temperature gradient leading to a distinct conducting zone devoid of any convective flows. The incorporation of a stably stratified layer above $z = h$ breaks the mid-plane symmetry leading to the confinement of otherwise quasi-geostrophic columnar convection rolls in the $z$ direction.

The convection rolls are relatively thicker under slow rotation ($E = 10^{-3}$) [figs. \ref{f_nonmag_contour}a -d] than the corresponding rapid rotation cases at ($E = 10^{-4}$) [Figs. \ref{f_nonmag_contour}e - h]. At low rotation rates stationary onset modes occur for all values $Pr = 0.01,0.1,1$ whereas oscillatory convection occurs at low $Pr=0.01$ only with increased rotational dominance ($E = 10^{-4}$). The transition from stationary to oscillatory onset with lowering of $Pr$  values for rapid rotation regimes \cite{garai2022convective} significantly affects the penetrative convection characteristics.

The lateral localization of convective instabilities also gets weaker at such low $Pr=0.01$ regimes (Fig \ref{f_nonmag_contour}c,d) which is further homogenized due to the rapid rotation-induced oscillatory onset at $E = 10^{-4}$ (Fig \ref{f_nonmag_contour}g,h). In contrast, for stationary convection at $Pr = 1$, rotational dominance favors the clustering of the thinner columnar rolls (Fig \ref{f_nonmag_contour}e,f). 

\begin{figure}[htbp]
    \centering
    \includegraphics[clip, trim = 0cm 22cm 0cm 0cm, width=1\textwidth]{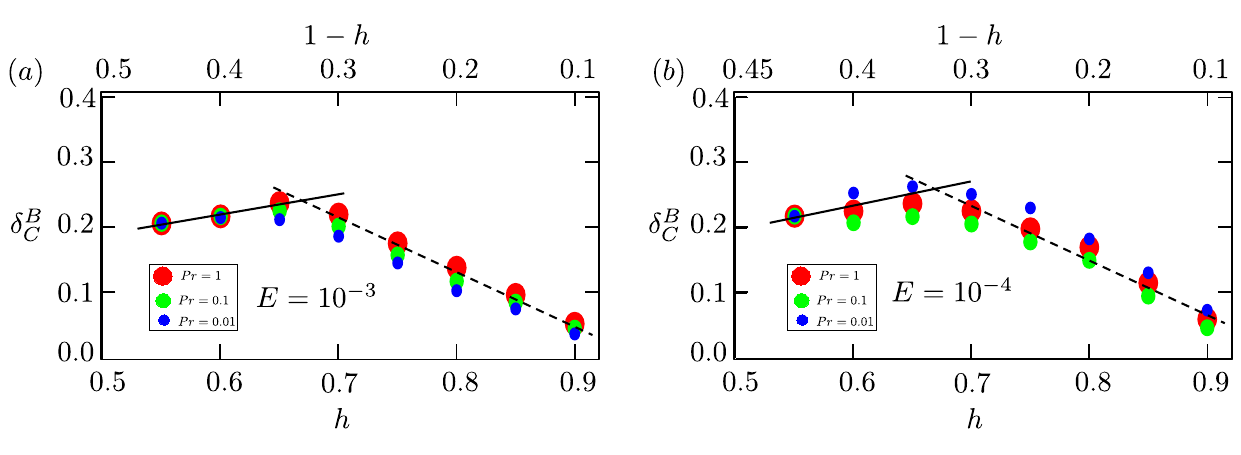}
    \caption{$\delta_{C}^B$, $h$, $1-h$ represent the penetration depth derived from the direct numerical simulations (DNS), the interface height, and the thickness of the stable layer, respectively. (a) and (b) represent cases of moderate rotation ($E = 10^{-3}$) and rapid rotation ($E = 10^{-4}$), respectively. Filled circles indicate different Prandtl numbers such as red(large) for $Pr = 1$, green(moderate) for $Pr = 0.1$, and blue(small) for $Pr = 0.01$, respectively.}
    \label{f_nonmag_depth_numerical_non}
\end{figure}

In the plane layer configuration, at the convective onset, the penetration depth ($\delta_{N}$) of axial flow ($u_z^{\prime}$) instabilities with respect to the stable-unstable stratification interface ($z = h$) is estimated as,
\begin{equation} \label{eq_drop_u1}
\delta_{N} = z^* - h.
\end{equation}
Here, $z^*$ is the axial location such that the interface velocity has been damped by a factor $1/e$, i.e., approximately by $37\%$, given by
\begin{equation} \label{eq_drop_u2}
u_z^{\prime}|_{z=z^*} = \frac{1}{e} u_z^{\prime}|_{z=h},
\end{equation}
where $e \approx 2.718$ is the base of the natural logarithm. The difference between the axial location $z^*$ and $h$ is considered as the depth of penetration of convective flows obtained from direct numerical simulations. The depth of penetration ($\delta_{C}^B$) (is equal to $\delta_N$ for thermal convection, where superscript ($B$) and subscript ($C$) indicate the bounded fluid domain and thermal convection, respectively) for moderate and rapid rotation is shown in figure (\ref{f_nonmag_depth_numerical_non}) for various interface heights ($h$). It is observed that $\delta_{C}^B$ transitions from an increasing trend to a decreasing behavior with $h$ beyond a critical interface height ($h^* \approx 0.65$). For $h<h^*$, $\delta_{C}^B \approx 0.31h$ whereas for $h>h^*$, $\delta_{C}^B \approx -0.83h$. This existence of critical interface height indicates the influence of the impenetrable top boundary on $\delta_{C}^B$ which, otherwise, would have retained its increasing trend for all values of $h$. The expected penetration depth estimates are indicated by a solid black line beyond $h = h^*$. Note that for $h \rightarrow 1$ the stable layer thickness $(1-h)$ clearly falls below the expected penetration depth estimates [Figs.\ref{f_nonmag_depth_numerical_non} a,b] indicating lack of sufficient axial range for avoiding the influence of the top boundary. 

The role of viscous fluid properties such as viscous and thermal diffusivities in modulating the penetration depth is investigated further in terms of the sensitivity of $\delta_{C}^B$ to $Pr$. The $Pr$ sensitivity $(\Delta \delta_{C}^B(h))$ is estimated as, 
\begin{equation} \label{delta}
\Delta \delta_{C}^B(h) = \delta_{C}^B|_{max} - \delta_{C}^B|_{min},   
\end{equation}
which is the difference between the maximum and minimum penetration depth for a given interface height ($h$). $\Delta \delta_{C}^B$ is maximum near $h = h^*$ and weakens away from the critical interface height. This is evident by nearly overlapping dots at $h = 0.55$ and $h = 0.9$, whereas a clear separation is present for $h = 0.65$ cases irrespective of $E$ [Figs \ref{f_nonmag_depth_numerical_non}a,b]. Quantitatively, the sensitivity ranges from $\Delta \delta_{C}^B \approx 10^{-4}$ at $h = 0.55, 0.9$ to $\delta_{C}^B = 0.03$ at $h = 0.7$. Although not realizable for the axially bounded fluid domain in-plane layer configuration, it is reasonable to expect that the reduction in sensitivity beyond $h = 0.7$ is due to the influence of the impenetrable top boundary. It is thus anticipated that in the absence of the top boundary, $\delta_{C}^B$ would have increased monotonically. 

The effect of enhanced rotation rates is nearly negligible for penetrative convection occurring as stationary modes as evidenced by similar $\delta_{C}^B$ values for $Pr = 1, 0.1$ regimes [Figs.\ref{f_nonmag_depth_numerical_non}b]. However, at the lowest value of $Pr = 0.01$, the penetration depth is significantly higher for the oscillatory onset mode at $E = 10^{-4}$ compared to the stationary mode at $E = 10^{-3}$. Such enhancement in penetration depth may be attributed to the time-varying aspect of the onset mode instabilities. 

\begin{figure}[htbp]
    \centering
    \includegraphics[clip, trim = 0cm 22cm 0cm 0cm, width=1\textwidth]{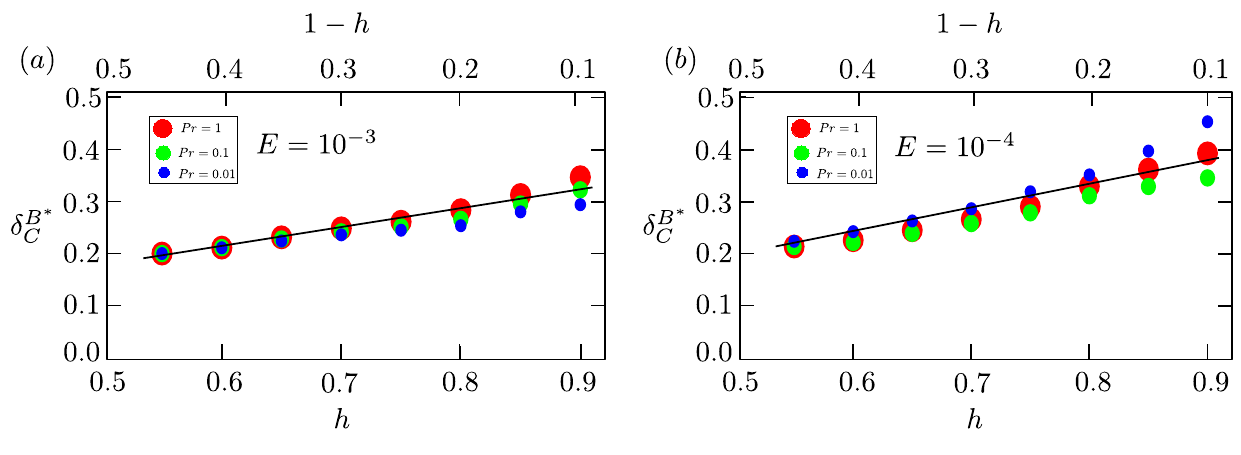}
    \caption{$\delta_{C}^{B^*}$, $h$, $1-h$ represent the normalized depth of penetration, the interface height, and the thickness of the stable layer respectively. (a) and (b) represent moderate rotation ($E = 10^{-3}$) and rapid rotation ($E = 10^{-4}$), respectively. Filled circles indicate different Prandtl numbers such as red(large) for $Pr = 1$, green(moderate) for $Pr = 0.1$, and blue(small) for $Pr = 0.01$, respectively.}
    \label{f_nonmag_depth_numerical}
\end{figure}

The confinement effects of the impenetrable top boundary can plausibly be applicable to the Earth's outer core conditions where the $CMB$ prevents any mass transfer from the outer core to the lower mantle. However, given the large uncertainties in the thickness of the Earth's outer core stable layer \cite{buffett2010stratification, helffrich2010outer, gubbins2015core, greenwood2021evolution}, the underlying convective perturbation may not be affected by the $CMB$. Moreover, regionally enhanced stratification \cite{mound2019regional} may be more than sufficient for $CMB$ confinement effects to affect the penetration of convective flows. Also, the possible existence of stable stratification in gas giants without any confining boundaries \cite{dietrich2018penetrative, moore2022dynamo}, may be better addressed with the present model without the confinement effects. Hence, an attempt is made to characterize the confinement effects and estimate the penetration depth without such effects.

The confinement effects on the penetration depth ($\delta_{C}^B$) is estimated by adapting equation (\ref{eq_drop_u1}) in terms of the thickness of stable layer $(1-h)$ as,
\begin{equation} \label{eq_delta_numerical}
\delta_{C}^{B^*} = \frac{\delta_{C}^B}{\alpha (1-h)^{\gamma}} + \epsilon
\end{equation}
The model parameters ($\alpha$, $\gamma$, and  $\epsilon$) are estimated based on the following three constraints and $^*$ in superscript in $\delta_C^{B^*}$ indicates normalization. Firstly, the values of $\delta_{C}^{B}$ for $h < h^{\star}$. Additionally, the slope and intercept of the best fit straight line are unchanged for both $\delta_{C}^{B}$ and $\delta_{C}^{B^*}$. Based on the above constraints, the value of the exponent in the denominator of equation (\ref{eq_delta_numerical}) is obtained as $\gamma = 2$. The values of ($\alpha$, $\epsilon$) are obtained as (29.63, 0.166) and (27.57, 0.175) for $E = 10^{-3}$ and $E = 10^{-4}$ regimes respectively. Thus, the confinement effects are quantitatively characterized by the corresponding values of ($\alpha$, $\gamma$, and $\epsilon$) such that a higher value of $\alpha$ indicates a larger confinement effect. The estimates of penetrative depths, after elimination of confinement effects ($\delta_{C}^{B^*}$), vary monotonously with $h$ for both lower (Fig.\ref{f_nonmag_depth_numerical}a) and higher rotational  (Fig.\ref{f_nonmag_depth_numerical}b) regimes. The parametric dependencies of the depth estimates with respect to $E$ and $Pr$ remain unchanged as imperative. The clear alignment of ($\delta_{C}^{B^*}$) over the entire $h$ range indicates that the confinement effect of the boundary is relaxed. 

\subsubsection{Penetrative magneto-convection}
The instability characteristics at the onset of convection, in the presence of a background magnetic field, depend on the orientation and strength of the imposed magnetic field. The presence of symmetry-breaking constraints such as stable stratification \cite{sreenivasan2024oscillatory} or in-homogeneous thermal boundary conditions \cite{sreenivasan2017confinement} has been shown to modify the magnetoconvection onset characteristics for homogeneous buoyancy forcing configurations. Going further, in the present study, the penetration depth of onset instabilities for heterogeneous magneto-convection (MC) configuration is investigated in the limit of strong rotational effects with either axially ($\mathbf{B}^* = B^* \hat{z}$, $AMC$) or horizontally ($\mathbf{B}^* = B^* \hat{x}$, $HMC$) oriented magnetic fields. These magnetic field orientations are motivated from the poloidal and toroidal magnetic fields, respectively, occurring in numerical geodynamo simulations \cite{christensen1999numerical} modeling the Earth's outer core dynamics.  

\begin{figure}[htbp!]
\centering
\includegraphics[clip, trim = 0cm 20cm 0cm 0cm, width=1\textwidth]{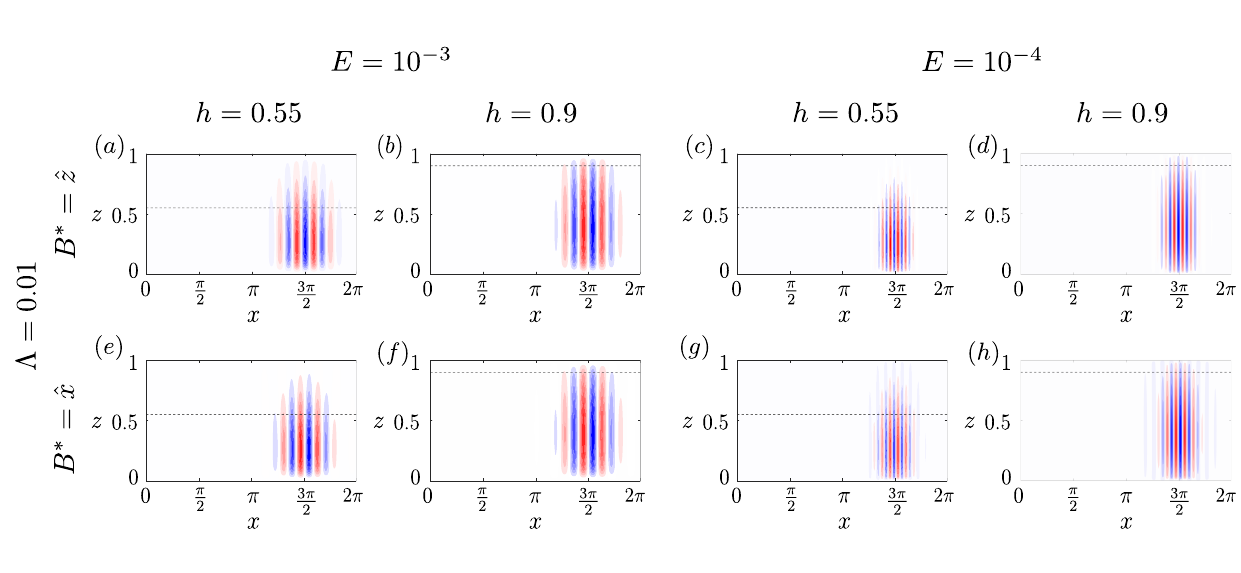}
\caption{(a)-(h) represent contour maps of axial velocity ($u_z^{\prime}$) at onset for rotationally dominating regime at $\Lambda = 0.01$ for penetrative magneto-convection. (a)-(d) indicate $AMC$ for vertically imposed magnetic field ($B^* = \hat{z}$) and (e)-(h) indicate $HMC$ for horizontally imposed magnetic field ($B^* = \hat{x}$). (a,e) for transition height at $h = 0.55$, (b,f) for transition height at $h = 0.9$ for moderate rotation ($E = 10^{-3}$), and (c,g) for transition height at $h = 0.55$, (d,h) for transition height at $h = 0.9$ for rapid rotation ($E = 10^{-4}$).}
\label{f_contour_low_mag}
\end{figure}

The onset of convection have been obtained using numerical simulations for the rotationally dominating regime represented by a low value of $\Lambda = 0.01$. The axial velocity ($u_z^{\prime}$) contours for two values of $h$ denoting thick ($h = 0.55$) and thin ($h = 0.9$) stably stratified layers clearly indicate the corresponding penetrative effects with respect to the stable-unstable interface indicated by the dotted line [Figs. (\ref{f_contour_low_mag})]. Such penetrative effect of convective instabilities at onset is robustly observed irrespective of changes in the rotation rates or the orientations of the magnetic field (compare Figs. \ref{f_contour_low_mag} a,b,e,f with \ref{f_contour_low_mag} c,d,g,h, respectively). On the other extreme, the penetration depth of convective flows, in the limit of magnetic dominance represented by a high value of $\Lambda = 5$, with either axially ($\mathbf{B}^* = B^* \hat{z}$) or horizontally ($\mathbf{B}^* = B^* \hat{x}$) oriented magnetic fields, is also explored. 

Compared to the rotationally dominating regime [Figs. \ref{f_contour_low_mag}], $AMC$ exhibits slightly thicker columnar flow structures [Fig.(\ref{f_contour_high_mag} a - d)] while the convection rolls occurring at onset for $HMC$ configurations have significantly larger horizontal wavelength [Figs. (\ref{f_contour_high_mag} e - h)]. This indicates a relatively weaker influence of the axial fields in promoting large scales in rotating convection while horizontal fields have a much stronger effect. The inherent effects of rapid rotation in reducing the lateral length scales and clustering of columnar rolls in the region of enhanced temperature gradients remain unaffected.  

\begin{figure}[htbp!]
    \centering
    \includegraphics[clip, trim = 0cm 20cm 0cm 0cm, width=1\textwidth]{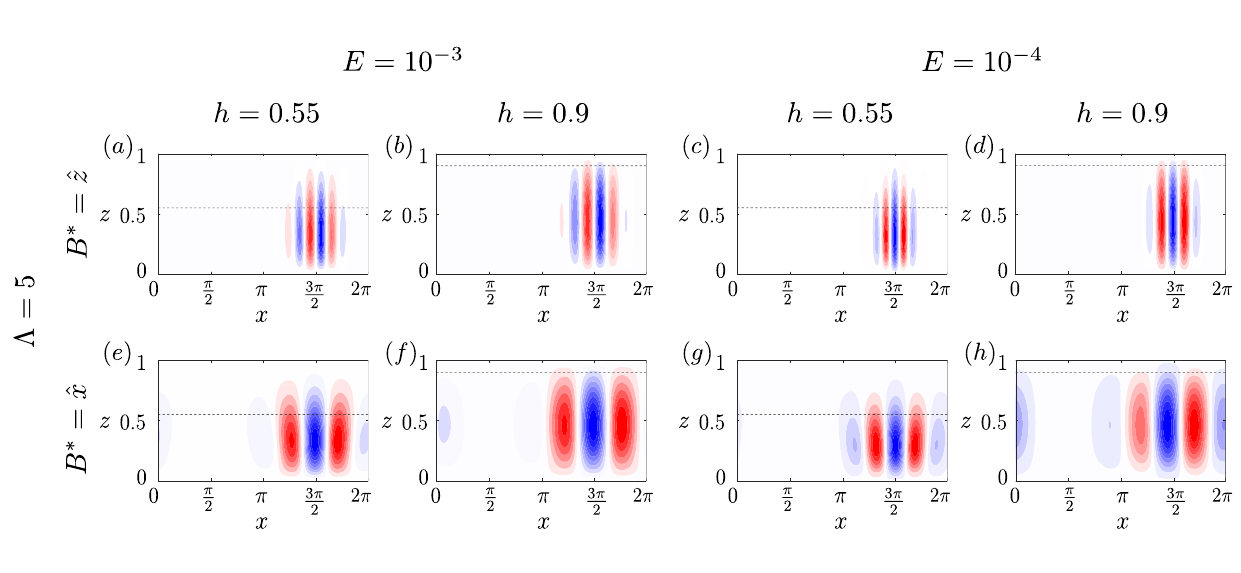}
    \caption{(a)-(h) represent contour maps of axial velocity ($u_z^{\prime}$) at onset for magnetically dominating regime at $\Lambda = 5$ for penetrative magneto-convection. (a)-(d) indicate $AMC$ for vertically imposed magnetic field ($B^* = \hat{z}$) and (e)-(h) indicate $HMC$ for horizontally imposed magnetic field ($B^* = \hat{x}$). (a,e) for transition height at $h = 0.55$, (b,f) for transition height at $h = 0.9$ for moderate rotation ($E = 10^{-3}$), and (c,g) for transition height at $h = 0.55$, (d,h) for transition height at $h = 0.9$ for rapid rotation ($E = 10^{-4}$).}
    \label{f_contour_high_mag}
\end{figure}

The quantitative estimation for the penetration depth of convection rolls is performed for the above configurations using equations (\ref{eq_drop_u1}) and (\ref{eq_drop_u2}). As the value of $z^*$ depends on the horizontal location ($x$) due to lateral temperature variations, the location of maximum temperature gradient ($x = \frac{3 \pi}{2}$) is chosen for estimating the penetration depth ($\delta_{MC}^{B}$) (is equal to $\delta_N$ for magnetoconvection, where superscript ($B$) and subscript ($MC$) indicate the bounded fluid domain and magnetoconvection convection, respectively). The estimates have been performed for both rotationally and magnetically dominating regimes for various thicknesses of the stable layer ($1-h$).

\begin{figure}[htbp]
    \centering
    \includegraphics[clip, trim = 0cm 22cm 0cm 0cm, width=1\textwidth]{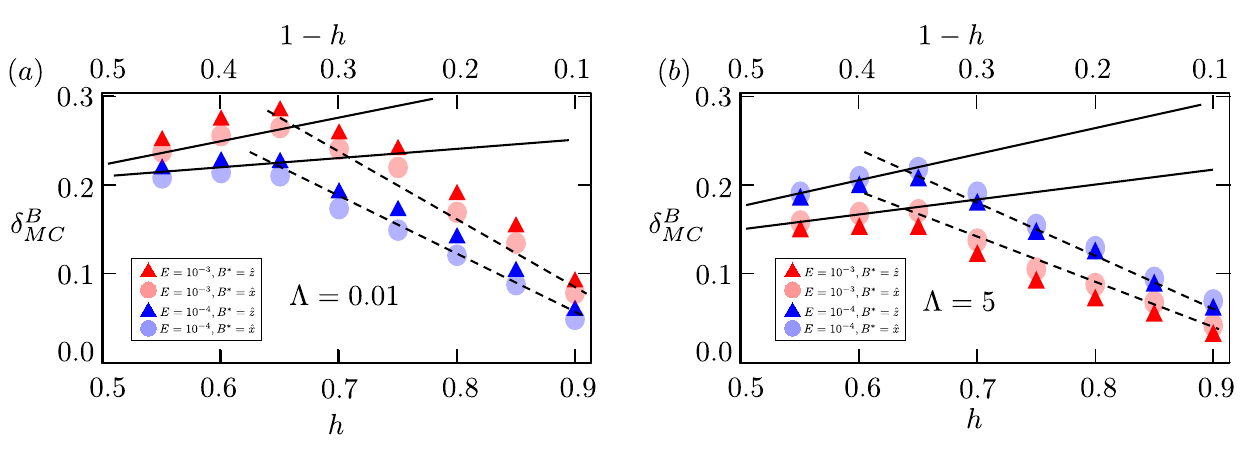}
    \caption{$\delta_{MC}^{B}$, $h$, $1-h$ represent the penetration depth derived from the direct numerical simulations (DNS), the interface height, and the thickness of the stable layer respectively. (a) and (b) represent rotationally dominating regime ($\Lambda = 0.01$) and  magnetically dominating regime ($\Lambda = 5$), respectively. Filled triangles indicate $AMC$ ($B^* = \hat{z}$) and filled circles indicate $HMC$ ($B^* = \hat{x}$). Red color represent moderate rotation ($E = 10^{-3}$) and blue color represent rapid rotation ($E = 10^{-4}$), respectively.}
\label{f_mag_depth_numerical_non}
\end{figure}

Qualitatively, even for magnetoconvection, $\delta_{MC}^{B}$ varies with $h$ in a manner similar to the purely thermal convection with the critical interface height at $h^* \approx 0.65$ (Fig \ref{f_mag_depth_numerical_non}).
For the cases with rotational dominance ($\Lambda = 0.01$, [Fig \ref{f_mag_depth_numerical_non}a]), $\delta_{MC}^{B} \approx +0.46 h$  at $E = 10^{-3}$ and $\delta_{MC}^{B} \approx +0.17 h$ at $E = 10^{-4}$ for $h < h^*$. Beyond the critical interface height, confinement effects due to the top boundary lead to negative slopes for $\delta_{MC}^{B}-h$ variation. Thus, for $h > h^*$, $\delta_{MC}^{B} \approx -0.83h$ at $E = 10^{-3}$ and $\delta_{MC}^{B} \approx -0.71h$ at $E = 10^{-4}$. 

The corresponding best-fit lines for the estimated values of penetrative depths for 
magnetically dominating regime ($\Lambda = 5$, [Fig \ref{f_mag_depth_numerical_non}b])
indicate an almost invariance at $E = 10^{-3}$ while significant reduction in $\delta_{MC}^{B}$ at $E = 10^{-4}$. The variation of $\delta_{MC}^{B}$ becomes stronger with $h$ as indicated by enhance slopes with magnetic dominance.  Quantitatively, for  $h < h^*$, $\delta_{MC}^{B} \approx +0.53h$ at $E = 10^{-3}$ and $\delta_{MC}^{B} \approx +0.27h$ at $E = 10^{-4}$  irrespective of the orientations of the magnetic field. The confinement effects negate the slopes for $h > h^*$ resulting in $\delta_{MC}^{B} \approx -0.45h$ at $E = 10^{-3}$ and $\delta_{MC}^{B} \approx -0.61h$ at $E = 10^{-4}$. 

Overall, it is observed that in the presence of strong (weak) background magnetic fields, the penetration depths become enhanced (reduced) with rotational dominance. Such 
transition of the penetration depth results from a drastic reduction in $\delta_{MC}^{B}$ for lower $E$ cases while the depth remains almost invariant at $E= 10^{-3}$. This may be explained by the lower sensitivity of penetration depth to $\Lambda$ at $E = 10^{-4}$ 
due to stronger axial invariance. In most cases, the lower magnitude of the slopes at $E = 10^{-4}$ indicates reduced sensitivity of the penetration depth to the stable layer thickness. This may be attributed to the enhanced axial invariance due to stronger geostrophy for rapidly rotating regimes which acts against the axial damping. The change in the orientations of the imposed magnetic field seems to have a negligible effect on the penetration depth. The influence of the confining upper boundary is found to be a robust effect, evident by the divergent solid and dotted lines representing the expected and actual penetration depths.

Unlike, thermal convection, the presence of even a weak magnetic field ($\Lambda = 0.01$) leads to considerable rotational dependence of $\delta_{MC}^{B}$. The change in the penetration depth is maximum ($0.12$) at $h = 0.7$, close to the critical interface height $h^*$. Such dependency on $E$ remains unaffected even for strong magnetoconvection at $\Lambda = 5$. Interestingly, the penetration depth for $AMC$ is marginally higher compared to that with $HMC$ for rotationally dominant regimes. Although quantitatively small, this trend is reversed for strong field magnetoconvection.

Furthermore, the confinement effects on the penetration depth ($\delta_{MC}^{B}$) are estimated using equation (\ref{eq_delta_numerical}), in a manner similar to the case of non-magnetic convection. The value of the $\gamma = 2$ is preserved even for magnetoconvection configuration indicating that the confinement effects are fundamentally independent of the imposition of magnetic fields and are solely determined based on the stable layer thickness $1-h$. For the rotationally dominant regime, the values of ($\alpha$, $\epsilon$) are obtained as (24.8, 0.191) and (63.97, 0.221) for $E = 10^{-3}$ and $E = 10^{-4}$ regimes respectively [Figs. \ref{f_mag_depth_numerical}a]. For magnetically dominant regime, the values of ($\alpha$, $\epsilon$) are obtained as (29.80, 0.141) and (25.8, 0.154) for $E = 10^{-3}$ and $E = 10^{-4}$ regimes respectively [Figs. \ref{f_mag_depth_numerical}b]. The above quantification of the confinement effect is applicable to both $AMC$ and $HMC$, strengthening the assertion that such effects do not depend on background magnetic fields. In line with the purely thermal convection system, the parametric dependencies of the depth estimates with respect to $E$ and $Pr$ remain unchanged even for magnetoconvection. The resulting alignment of ($\delta_{MC}^{B^*}$) over the entire $h$ range indicates that the confinement effect of the boundary is relaxed. For all the cases, the adapting parameter $\epsilon \sim 10^{-1}$ remains limited to values much less than unity. 

\begin{figure}[!]
\centering
\includegraphics[clip, trim = 0cm 23cm 0cm 0cm, width=1\textwidth]{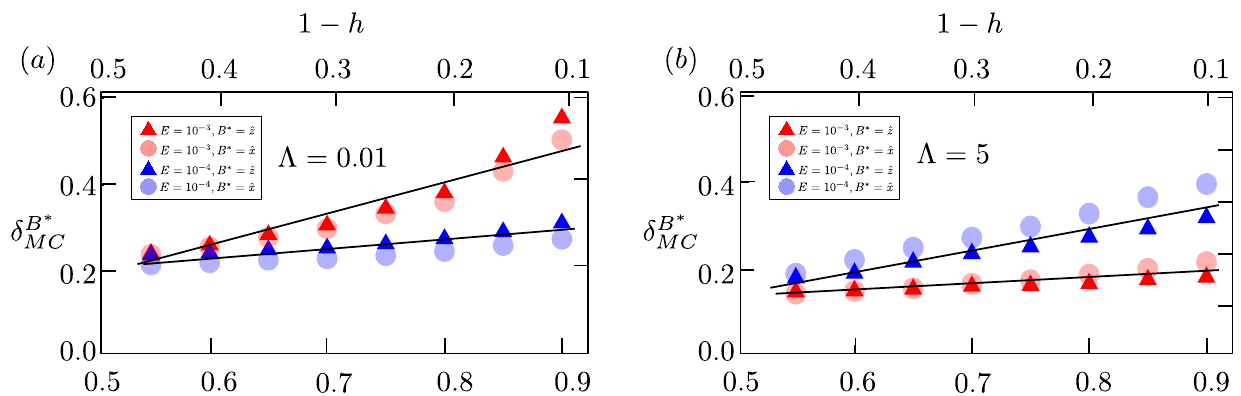}
\caption{$\delta_{MC}^{B^*}$, $h$, $1-h$ represent the normalized depth of penetration, the interface height, and the thickness of the stable layer respectively. (a) and (b) represent rotationally dominating regime ($\Lambda = 0.01$) and  magnetically dominating regime ($\Lambda = 5$), respectively. Filled triangles indicate $AMC$ ($B^* = \hat{z}$) and filled circles indicate $HMC$ ($B^* = \hat{x}$). Red color represent moderate rotation ($E = 10^{-3}$) and blue color represent rapid rotation ($E = 10^{-4}$), respectively.}
\label{f_mag_depth_numerical}
\end{figure}

\subsection{Penetrative convection in unbounded fluid}
Penetrative convective flows driven by buoyancy forcing in unbounded fluid domain offer a simple mathematical problem amenable to theoretical analysis, providing significantly useful estimates of the depth of penetration as close-form expressions. The role of laterally heterogeneous buoyancy, in the presence of stable stratification, is investigated below using a bi-global analysis approach. The penetrative action of the stable stratification on convective flows driven by rotating thermal convection with and without the presence of background magnetic fields is formulated. The analysis is performed for physically insightful and relevant buoyancy configurations, although simplified compared to planetary core conditions in the interest of obtaining exact closed-form solutions.

\subsubsection{Penetrative thermal convection}
Localized or non-localized disturbances in an incompressible fluid under uniform rotation can propagate as waves, known as inertial waves \cite{davidson2014dynamics}.
Thermal forcing due to axial buoyancy results in Inertial-Gravity (IG) waves sustainable in stably stratified media\cite{maurer2016generation, mukherjee2023thermal}. Additional imposition of laterally varying temperature gradients can lead to substantial modifications to the wave characteristics hereby termed as modified inertial-gravity (mIG) wave.    
 
In the present study, the depth of penetration of mIG wave in an unbounded fluid domain with partial thermally stable stratification is considered. To determine the characteristics of an undamped $mIG$ wave, consider the equations (\ref{eq_incom_dimensionless_perturb}) - (\ref{eq_temp_dimensionless_perturb}) with the non-linear, viscous, and thermal diffusion terms neglected. After global normal mode analysis, the dispersion relation is obtained (Appendix A) providing the exact closed-form expression for the frequency as,
\begin{equation} \label{eq_dispersion_hetero_full_3ab}
\zeta = \pm \left[  \left ( \frac{Pr}{E} \right)^2 \frac{m^2}{k^2 + m^2} + RaPr \Gamma_z^* \frac{{k}^2}{k^2 + m^2} -  RaPr \Gamma_x^* \frac{k m}{k^2 + m^2} \right]^{1/2}.
\end{equation}
where, $k, m$ are the wavenumbers in $x, z$ directions respectively.

Inherently, the frequency of $mIG$ waves depends on the wavenumbers indicating 
dispersive and anisotropic characteristics. The additional second and third terms on the RHS denote the effects of stable stratification and lateral buoyancy heterogeneity respectively. Note that, in absence of stable stratification ($\Gamma_z^{\star} = 0$) and lateral variations ($\Gamma_x^{\star} = 0$). the frequency of purely inertial waves is obtained \cite{davidson2014dynamics}.
Rearranging the terms in equation (\ref{eq_dispersion_hetero_full_3ab}) to obtain closed form expression for $m$ results in,
\begin{equation} \label{eq_dispersion_hetero_full_rot_m}
m = \frac{-k RaPr \Gamma_x^* \mp i k \sqrt{4 \left ( \left ( \frac{Pr}{E} \right)^2 - \zeta^2 \right) (RaPr \Gamma_z^* - \zeta^2) - (RaPr \Gamma_x^*})^2}{2( \left ( \frac{Pr}{E} \right)^2 - \zeta^2)}
\end{equation}
Consider the regime where the rotational time scale ($\propto Pr/E$) is much larger than $mIG$ wave oscillation period ($\propto 1/\zeta$). Thus, assuming $\frac{Pr}{E} >> \zeta$ 
and $RaPr\Gamma_z^* >> \zeta$ denoting the limit of strong stable stratification\cite{takehiro2001penetration}, $m$ is simplified to,
\begin{equation} \label{eq_dispersion_hetero_full_rot_mod}
m = \frac{- k RaPr \Gamma_x^* \mp i k \sqrt{4 \left( \frac{Pr}{E} \right)^2 Ra Pr \Gamma_z^* - (RaPr \Gamma_x^*})^2}{2\left( \frac{Pr}{E} \right)^2 }
\end{equation}

The penetrative length scale ($\delta^{U}_{C}$, superscript ($U$) and subscript ($C$) indicate unbounded fluid and thermal convection, respectively) of mIG waves can be obtained as the inverse of the positive imaginary part in the above solution ($\delta^{U}_{C} = |Imag[m]|^{-1}$) \cite{takehiro2001penetration}. Thus the characteristic depth of penetration of the mIG wave in the axial direction is obtained as,
\begin{equation} \label{eq_penetration_depth_hetero}
\delta^{U}_{C} = \left (\frac{2 \left ( \frac{Pr}{E} \right)^2}{\sqrt{4 \left( \frac{Pr}{E} \right)^2 Ra Pr \Gamma_z^* - (RaPr \Gamma_x^*)^2}}\right) \frac{1}{k}
\end{equation}
The depth of penetration $\delta^{U}_{C}$ is proportional to the horizontal length scale ($\frac{1}{k}$) and $Pr$ values. At the same time, it is inversely proportional to $E, Ra, \Gamma_z^*, \Gamma_x^*$. An increase in axial temperature gradient ($\Gamma_z^*$) would result in a reduction in $\delta^{U}_{C}$, while enhancement in $\Gamma_x^*$ would result in a further decrease in penetration depth. For homogeneous horizontal temperature gradient ($\Gamma_x^* = 0$), the depth of penetration (\ref{eq_penetration_depth_hetero}) of unmodified $IG$ waves simplifies to,
\begin{equation} \label{eq_penetration_depth_homo}
\delta^{hom}_{C} = \left (\frac{ \frac{Pr}{E} }{\sqrt{ Ra Pr \Gamma_z^*}}\right) \frac{1}{k}
\end{equation}
Note that upon dimensionalization of the above equation, the depth of penetration for reference homogeneous cases is obtained as
\begin{equation} \label{eq_penetration_depth_homo_dimensional_nonmag}
\delta^{hom}_{C} = \frac{2 \Omega}{N_z}\frac{1}{k}
\end{equation}
as obtained in previous studies \cite{takehiro2001penetration}. Superscript ($hom$) in both the above equations (\ref{eq_penetration_depth_homo} and \ref{eq_penetration_depth_homo_dimensional_nonmag}) indicate without laterally heterogeneous thermal gradient case.

\begin{figure}[htbp]
\centering
\includegraphics[clip, trim = 0cm 22cm 0cm 0cm, width=1\textwidth]{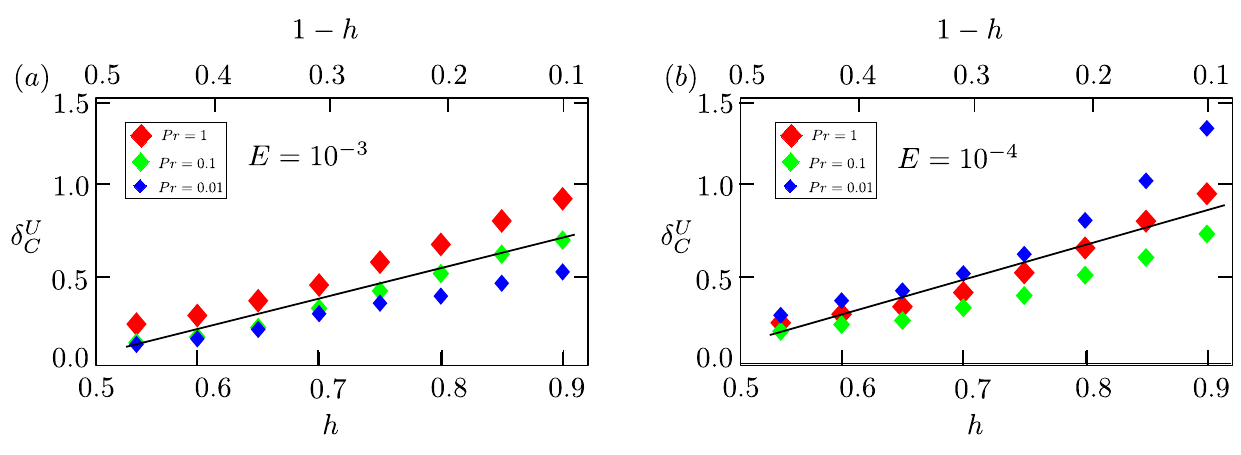}
\caption{$\delta^{U}_{C}$, $h$, $1-h$ represent the depth of penetration estimated from analytical expression, the interface height, and the thickness of the stable layer respectively. (a) and (b) represent moderate rotation ($E = 10^{-3}$) and rapid rotation ($E = 10^{-4}$), respectively. Filled circles indicate different Prandtl numbers such as red(large) for $Pr = 1$, green(moderate) for $Pr = 0.5$, and blue(small) for $Pr = 0.01$, respectively.}
\label{f_nonmag_depth_analytical}
\end{figure}

The penetration depth for various interface heights of stratification ($h$), calculated using equations (\ref{eq_penetration_depth_hetero}) and (\ref{eq_gammaZ-X}), indicates a monotonous enhancement for both weakly rotating [Figs. (\ref{f_nonmag_depth_analytical}a)] and rapidly rotating [Figs. (\ref{f_nonmag_depth_analytical}b)] regimes and $Pr$ values. In line with numerical simulations for penetrative convection, the estimates for penetration depth in the unbounded domain remain sensitive to $Pr$, especially for higher values of $h$. Moreover, the parametric dependencies also remain unchanged from those obtained for the plane layer model. Interestingly, the monotonic trend observed for $\delta^T$ with the unbounded domain and the linear alignment of the depth values conforms to that obtained for the plane layer model after the elimination of the confinement effects. Also, the enhanced sensitivity of $\delta^{U}_{C}$ to $Pr$ for higher $h$ values is qualitatively similar to the estimates for $\delta_{MC}^{B^*}$ after removal of confinement effects (compare Fig.\ref{f_nonmag_depth_analytical} with Fig.\ref{f_nonmag_depth_numerical}). 

\subsubsection{Penetrative magneto-convection}
Including a background magnetic field imparts rigidity to the fluid in the unbounded domain. Consequently, the MHD disturbance in that fluid can propagate along the imposed magnetic field as waves, known as Alfv\'en waves \cite{finlay2008course}. The incorporation of uniform rotation leads to magnetic-Coriolis (MC) waves \cite{finlay2008course} by the combined effect of magnetic field and vortex tension. Further incorporation of axial thermal buoyancy results in Magnetic-Archimedes-Coriolis (MAC) waves \cite{finlay2008course}. However, the additional impact of laterally varying thermal heterogeneity can modify the wave characteristics substantially, which has been delineated in this section. Hence, the depth of penetration of the modified Magnetic-Archimedes-Coriolis (mMAC) waves in an unbounded fluid is derived.

Consider the rotationally dominating regimes ($\Lambda << 1$) with the Coriolis force is much stronger than Lorentz forces. This results in magnetically modified internal gravity ($mmIG$) waves which are closer to internal gravity waves modified by lateral heterogeneous thermal gradients and weakly affected by background magnetic fields. The dispersion relation for the $mmIG$ waves (derived in Appendix B: Eqs.   ($\ref{eq_gravity_wave_dispersion_hetero_full_rot}$)) is given by,
\begin{equation} \label{eq_dispersion_hetero_full_3b}
\zeta = \pm \left[  \left ( \frac{Pm}{E} \right)^2 \frac{m^2}{k^2 + m^2} + qRaPm \Gamma_z^* \frac{{k}^2}{k^2 + m^2} -  qRaPm \Gamma_x^* \frac{k m}{k^2 + m^2} \right]^{1/2}.
\end{equation}
Solving the above equation (\ref{eq_dispersion_hetero_full_3b}) for $m$ results in,
\begin{equation} \label{eq_gravity_wave_dispersion_hetero_full_rot_m}
m = \frac{-k qRaPr \Gamma_x^* \mp i k \sqrt{4 \left ( \left ( \frac{Pm}{E} \right)^2 - \zeta^2 \right) (qRaPm \Gamma_z^* - \zeta^2) - (qRaPm \Gamma_x^*})^2}{2( \left ( \frac{Pm}{E} \right)^2 - \zeta^2)}
\end{equation}
Further assumptions due to the asymptotic regimes of rapid rotation ($\frac{Pm}{E} >> \zeta$) and strong stratification ($qRaPm\Gamma_z^* >> \zeta$), equation (\ref{eq_gravity_wave_dispersion_hetero_full_rot_m}) becomes,
\begin{equation} \label{eq_gravity_wave_dispersion_hetero_full_rot_mod}
m = \frac{- k qRaPm \Gamma_x^* \mp i k \sqrt{4 \left( \frac{Pm}{E} \right)^2 qRa Pm \Gamma_z^* - (qRaPm \Gamma_x^*})^2}{2\left( \frac{Pm}{E} \right)^2}
\end{equation}
In line with the procedure to obtain equation (\ref{eq_penetration_depth_hetero}), the positive imaginary part of the wavenumber $m$ in the above equation (\ref{eq_gravity_wave_dispersion_hetero_full_rot_mod}) is used to obtain the depth of penetration ($\delta^{U}_{MC}$) as,
\begin{equation} \label{eq_gravity_wave_penetration_depth_hetero}
\delta^{U}_{MC} = \left (\frac{2 \left ( \frac{Pm}{E} \right)^2}{\sqrt{4 \left( \frac{Pm}{E} \right)^2 qRa Pm \Gamma_z^* - (qRaPm \Gamma_x^*)^2}}\right) \frac{1}{k}
\end{equation}
In the case of vanishing lateral temperature gradients, setting $\Gamma_x^* = 0$ in the above equation (\ref{eq_gravity_wave_penetration_depth_hetero}), the depth of penetration is simplified to
\begin{equation} \label{eq_gravity_wave_penetration_depth_homo}
\delta^{hom}_{MC} = \left (\frac{ \frac{Pm}{E} }{\sqrt{ qRa Pm \Gamma_z^*}}\right) \frac{1}{k}
\end{equation}
Note that in terms of dimensional quantities, the depth of penetration for reference homogeneous cases is obtained from the above equation as
\begin{equation} \label{eq_penetration_depth_homo_dimensional_mag}
\delta^{hom}_{MC} = \frac{2 \Omega}{N_z}\frac{1}{k}
\end{equation}
as obtained in previous studies \cite{takehiro2001penetration}.

The corresponding dispersion relation expressing the frequency as a closed form dependency on wavenumber for the magnetically dominant regime ($\Lambda >> 1$) is given by (Appendix B, Equation \ref{eq_full_operator_MAC_dispersion})
\begin{equation} \label{eq_full_operator_MAC_dispersion_a}
\zeta = \pm \left[  \left ( \frac{\Lambda Pm}{E} \right)^2 \left(\mathbf{B}^* \cdot \mathbf{K} \right)^2 + qRaPr \Gamma_z^* \frac{{k}^2}{k^2 + m^2} -  qRaPr \Gamma_x^* \frac{km}{k^2 + m^2} \right]^{1/2}
\end{equation}
The fast waves ($\zeta > \zeta_A$) are filtered out by considering sufficiently small frequencies ($\zeta \rightarrow 0$)\cite{takehiro2015penetration}. Hence, the propagation and penetrative action of the retained slow waves \cite{xu2024penetrative, majumder2024self} are considered for both axial and horizontally aligned imposed magnetic fields.

For dimensionless vertically imposed magnetic field ($\mathbf{B}^* = \bm{\hat{z}}$), the slow waves propagate along the field lines in the axial direction ($\mathbf{B}^* \cdot \mathbf{K} = m$) resulting in,
\begin{equation} \label{eq_dispersion_vertical}
\left( \frac{\Lambda Pm}{E} \right) m^4 + \left( \frac{\Lambda Pm}{E} k^2 \right) m^2 + (- q Ra Pm \Gamma_x^* k ) m + qRa Pm \Gamma_z^* k^2 = 0
\end{equation}
Neglecting the higher order terms in the above equation (\ref{eq_dispersion_vertical}), as short waves with small wavenumber $m$ are unable to penetrate significantly \cite{xu2024penetrative}, the closed form solutions for $m$ are obtained as,
\begin{equation} \label{eq_dispersion_hetero_full_rot_vertical}
m = \frac{\left (qRaPm \Gamma_x^*\right) k \pm i \sqrt{4 q Ra Pm \Gamma_z^* \left( \frac{\Lambda Pm}{E} \right) k^4 - \left (q Ra Pm \Gamma_x^* k \right)^2}}{2 k^2 \left( \frac{\Lambda Pm}{E} \right)}
\end{equation}
The depth of penetration is estimated as the inverse of the positive imaginary part of the above equation (\ref{eq_dispersion_hetero_full_rot_vertical}) as,
\begin{equation} \label{eq_penetration_depth_hetero_vertical}
\delta^{U}_{MCa} = \frac{1}{ \sqrt{\left(\frac{qRaE }{\Lambda}\Gamma_z^* \right) - \left( \frac{q Ra E }{\Lambda} \Gamma_x^*\right)^2 \frac{1}{4 k^2}}}
\end{equation}
where, subscript $MCa$ indicates axial magnetoconvection ($AMC$).

In absence of laterally varying temperature gradient, by setting $\Gamma_x^* = 0$, in equation (\ref{eq_penetration_depth_hetero_vertical}), the depth of penetration is obtained as,
\begin{equation} \label{eq_penetration_depth_homo_vertical}
\delta_{MCa}^{hom} = \frac{1}{ \sqrt{\left(\frac{qRaE }{\Lambda}\Gamma_z^* \right)}}
\end{equation}
which is independent of the horizontal wave number ($k$). By considering $V_A = \frac{B_0^2}{\sqrt{\rho \mu}}$ is the pure Alfv\'en wave velocity, in dimensional form, the penetration depth becomes,
\begin{equation} \label{eq_penetration_Alfven_wave}
\delta_{MCa}^{hom} = \frac{V_A}{N_z}
\end{equation}

For horizontally imposed magnetic field ($\mathbf{B}^* = \bm{\hat{x}}$), the slow waves propagate along the field lines in the horizontal direction ($\mathbf{B}^* \cdot \mathbf{K} = k$) resulting in
\begin{equation} \label{eq_dispersion_horizontal}
\left( \frac{\Lambda Pm}{E} k^2 \right) m^2 + (- q Ra Pm \Gamma_x^* k ) m + \left (qRa Pm \Gamma_z^* k^2 + \frac{\Lambda Pm}{E} k^4 \right) = 0
\end{equation}
and the general solution for $m$ is given by,
\begin{equation} \label{eq_dispersion_hetero_full_rot_horizontal}
m = \frac{\left (qRaPm \Gamma_x^*\right) k \pm i \sqrt{4 (\frac{\Lambda Pm}{E}) \left (q Ra Pm \Gamma_z^* + \frac{\Lambda Pm}{E} k^2 \right ) k^4 - \left (q Ra Pm \Gamma_x^* k \right)^2}}{2 k^2 \left( \frac{\Lambda Pm}{E} \right)}
\end{equation}
The depth of penetration is estimated by considering the positive imaginary part of the above equation (\ref{eq_dispersion_hetero_full_rot_horizontal}) as,
\begin{equation} \label{eq_penetration_depth_hetero_horizontal}
\delta_{MCh}^{U} = \frac{1}{\sqrt{\left (\frac{qRaE }{\Lambda} \Gamma_z^* + k^2 \right) - \left( \frac{q Ra E }{\Lambda} \Gamma_x^* \right)^2\frac{1}{4 k^2}}}
\end{equation}
where, subscript $MCh$ indicates horizontal magnetoconvection ($HMC$). With laterally homogeneous temperature gradient ($\Gamma_x^* = 0$), the above equation (\ref{eq_penetration_depth_hetero_horizontal}) simplifies to 
\begin{equation} \label{eq_penetration_depth_homo_horizontal}
\delta_{MCh}^{hom} = \frac{1}{ \sqrt{\left(\frac{qRaE }{\Lambda}\Gamma_z^* + k^2 \right)}}
\end{equation}
Note that the depth of penetration extent of the slow wave depends on the horizontal wave number ($k$) which is unlike the case of the vertically imposed magnetic field (see Eqn. \ref{eq_penetration_depth_homo_vertical}).

\begin{figure}[htbp]
\centering
\includegraphics[clip, trim = 0cm 22cm 0cm 0cm, width=1\textwidth]{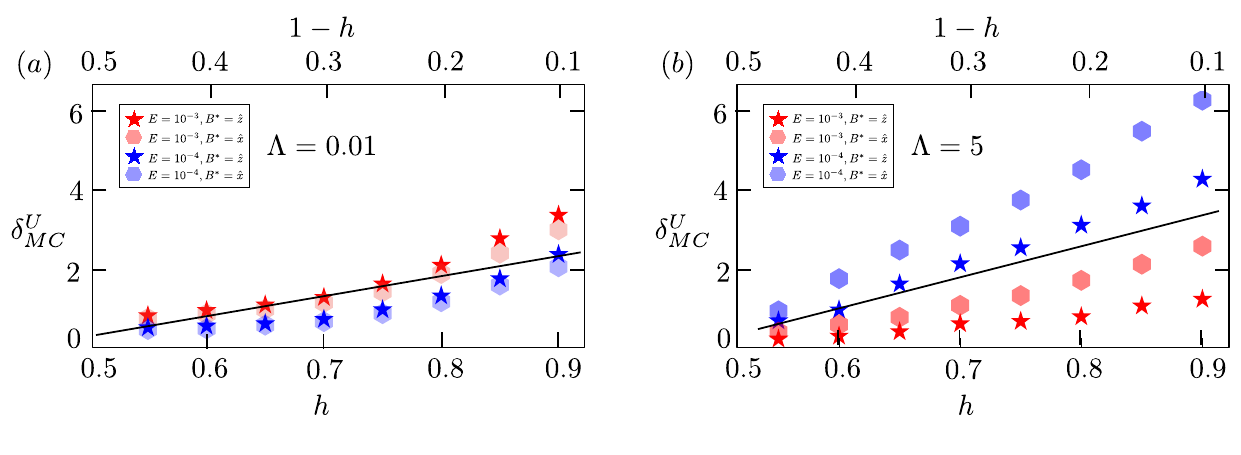}
\caption{$\delta^U_{MC}$, $h$, $1-h$ represent the depth of penetration estimated from analytical expression, the interface height, and the thickness of the stable layer, respectively. (a) and (b) represent rotationally dominating regime ($\Lambda = 0.01$) and  magnetically dominating regime ($\Lambda = 5$), respectively. Filled triangles indicate $AMC$ ($B^* = \hat{z}$) and filled circles indicate $HMC$ ($B^* = \hat{x}$). Red color represent moderate rotation ($E = 10^{-3}$) and blue color represent rapid rotation ($E = 10^{-4}$), respectively.}
\label{f_mag_depth_analytical}
\end{figure}

The calculated values for the penetration depth for both weak fields (Fig.\ref{f_mag_depth_analytical}a) and strong field  (Fig.\ref{f_mag_depth_analytical}b) regimes for various $h$ indicate a strong dependence on the strength of the imposed magnetic field. For weak magnetic fields, the values are evaluated using equation (\ref{eq_gravity_wave_penetration_depth_hetero}) and (\ref{eq_gammaZ-X}) for a fixed $Pr = Pm = 1$ regime, whereas, for strong magnetic fields with $AMC$ and $HMC$, the values are evaluated using equations (\ref{eq_penetration_depth_hetero_vertical}), (\ref{eq_penetration_depth_hetero_horizontal}), and (\ref{eq_gammaZ-X}), respectively. For weak imposed fields ($\Lambda = 0.01$), the penetration depth is much less sensitive to the directional orientation ($AMC/HMC$) as well as $E$. The influence of strong magnetic effects results in a clear separation of the $\delta^U_{MC}-h$ relation with rapid rotation leading to stronger penetration. Moreover, axial fields ($AMC$) allow slow waves-dominated and axially aligned columnar convective instabilities to penetrate deeper into the stably stratified regions compared to $HMC$ cases. Note that the monotonic $\delta^U_{MC}-h$ trends align with the penetration depth estimates ($\delta_{MC}^{B^*}$) obtained after the elimination of confinement effects (compare Figs. (\ref{f_mag_depth_analytical}) to Figs. (\ref{f_mag_depth_numerical})). 

\section{Discussion and Conclusions}\label{sec4}
The quantitative and qualitative characterization along with geometrical confinement effects on the extent of penetration of convection-driven flows
is the focus of the present study. The conditions in the Earth's outer core motivate the use of additional constraints such as background rotation, magnetic fields, and laterally varying temperature gradients which is the novelty of the present investigation. Direct numerical simulations and normal mode analysis have been utilized to obtain geometrically appropriate and conceptually insightful estimates of the penetration depth.

A novel hydrodynamic wave, designated as the modified inertial-gravity wave ($mIG$), arising due to the imposition of heterogeneous buoyancy is characterized using an augmented dispersion relation in the present study. Further imposition of background magnetic field leads to the magnetically modified inertial-gravity wave ($mmIG$) wave and magnetically modified gravity ($mmG$) in the rotationally dominated ($\Lambda << 1$) regime and magnetically dominated ($\Lambda >> 1$) regime, respectively.

From a geophysical perspective, penetrative convection in planetary liquid cores necessitates the use of appropriate buoyancy profiles with partially stable stratification. Simplified profiles with piecewise constant temperature gradients \cite{takehiro2001penetration, takehiro2015penetration,takehiro2018penetration, xu2024penetrative} is improved upon by the use of a non-linear temperature profile with gradients continuous across the stable-unstable interface in the present study. Interestingly, upon neglecting the influence of additional constraints of lateral heterogeneity, the generalized expression of penetration depth [Eqs.\ref{eq_penetration_depth_homo_dimensional_nonmag}] simplifies to that obtained in previous studies\cite{takehiro2001penetration} for non-magnetic thermal convection. 

The penetration depth of Aflv\'en waves along imposed axial magnetic fields has been estimated for strong field ($\Lambda >> 1$) regimes\cite{takehiro2015penetration} with negligible rotational effects. Such estimates of Aflv\'en wave penetration depth in magnetoconvection models, manifesting a direct proportionality to the speed and inverse proportionality to horizontal wave number ($k$), are further investigated to demonstrate a weaker dependence in geodynamo simulations\cite{gastine2020dynamo}. In the present study, both axial ($AMC$) and horizontal ($HMC$) fields are studied. For $AMC$, in the presence of lateral temperature gradients ($\Gamma_x^*$) the penetration depth of magnetically modified gravity ($mmG$) waves also depends on the horizontal wave number ($k$). However, for $\Gamma_x^* = 0$ the penetration depth [Eqs.\ref{eq_penetration_Alfven_wave}] becomes independent of $k$ and proportional to Alfv\'en wave velocity while inversely proportional to the stratification strength. For $HMC$, either in the presence [Eqs.\ref{eq_penetration_depth_hetero_horizontal}] or absence [Eqs.\ref{eq_penetration_depth_homo_horizontal}] of $\Gamma_x^*$, the penetration depth remains inversely proportional to $k$.  

The sensitivity of the penetration depth to the fluid properties is enhanced for stronger axial gradients in the stable stratification layer. Such dependence of penetrative depth on thermal and viscous diffusivities is a robust effect occurring even for spherical shell models of penetrative convection\cite{vidal2015quasi}. Furthermore, rotational effects lead to an overall decrease in the penetration depth for purely thermal convection with lateral buoyancy variations in line with previous studies based on linear stability analysis\cite{xu2024penetrative}. The transition from reduction to enhancement in penetration depth with stronger rotational effects is also observed for $mmIG$ waves in the presence of heterogeneous buoyancy. Such transition from rotational dominance to magnetic dominance for both axial and horizontal imposed fields is supported by recent studies\cite{xu2024penetrative}. Therefore, it can be anticipated that the penetration depth dependency on the convective length scale is strongly dependent on the choice of orientation of the magnetic field. With large-scale convective eddies, it may occur that the convective flows might not be able to penetrate more when the magnetic field is imposed in the axial direction in the absence of lateral heterogeneity whereas with a horizontally imposed magnetic field in the absence/presence of magnetic field the penetration of large scale convective eddies have higher possibilities. 

It is important to note a significant aspect of penetrative convection in plane layer models in the form of confinement effects. Characterized by a transition interface height separating the increasing and decreasing trends of penetration depth with stable layer thickness, such confinement effects are relaxed by appropriate normalization. The elimination of the confinement effect leads to a monotonic variation of penetrative depth with stratification strength displaying a good agreement with the trend estimated from the theoretically obtained closed-form expressions in the unbounded fluid domain. In the interest of generality, the applicability of the theoretical expressions, with suitable appropriations, to realistic spherical shell models of the outer core needs further investigation in future studies.

Penetrative convection and its consequent spatio-temporal variations in the flow behavior are strongly linked to geomagnetic secular variations for the Earth\cite{buffett2014geomagnetic}. Moreover, propositions of the plausible presence of thermally stably stratified layers in giant gaseous planets such as Jupiter\cite{moore2022dynamo, gastine2021stable} and the radiative zone of the Sun\cite{masada2013effects} enhance the applicability of the investigations performed in the present study. The inclusions of heterogeneous buoyancy and combinations of imposed magnetic fields further strengthen the pertinence to various other geophysical and astrophysical phenomena.

\section*{ACKNOWLEDGMENTS}
S.S. acknowledges the financial support through the award of the INSPIRE Faculty Fellowship by the Department of Science and Technology, India (Grant No. IFA18-EAS 70).

\section{APPENDICES}\label{sec5}
\appendix
\section{Penetrative thermal convection} 
Consider equations (\ref{eq_incom_dimensionless_perturb}) - (\ref{eq_temp_dimensionless_perturb}) which, after neglecting the non-linear, viscous, and thermal diffusion terms, 
give the linear perturbation equations as,
\begin{equation} \label{eq_incom_hetero}
\mathbf{\nabla} \cdot \mathbf{u^{\prime}} = 0,
\end{equation}
\begin{equation} \label{eq_NS_hetero}
\frac{\partial \mathbf{u}^{\prime}}{\partial t} + (\mathbf{u}^* \cdot \nabla) \mathbf{u}^{\prime} + (\mathbf{u}^{\prime} \cdot \nabla) \mathbf{u}^* + \frac{Pr}{E} (\hat{\textbf{z}} \times \mathbf{u}^{\prime}) = -\nabla P^{\prime} + Ra Pr T^{\prime}  \hat{\textbf{z}},
\end{equation}
\begin{equation} \label{eq_temp_hetero}
\frac{\partial T^{\prime}}{\partial t} + u_x^{\prime} \Gamma_x^* + u_z^{\prime} \Gamma_z^* = 0
\end{equation}
Operating $\hat{z} \cdot (\nabla \times)$ and $\hat{z} \cdot (\nabla \times \nabla \times)$ on equation (\ref{eq_NS_hetero}) results in,
\begin{equation} \label{eq_vor_hetero}
\frac{\partial \omega_z^{\prime}}{\partial t} = \frac{Pr}{E}(\hat{\textbf{z}} \cdot \nabla) u_z^{\prime}
\end{equation}
\begin{equation} \label{eq_velocity_hetero}
\frac{\partial }{\partial t} \nabla^2 u_z^{\prime} + \frac{Pr}{E}(\hat{\textbf{z}} \cdot \nabla) \omega_z^{\prime} = Ra Pr \nabla_h^2 T^{\prime}
\end{equation}
respectively, where, $\omega_z = \hat{z} \cdot (\nabla \times \mathbf{u})$ is the axial ($z$) component of vorticity and $\nabla^2_h = \nabla^2 - \frac{\partial^2}{\partial z^2}$ is the horizontal Laplacian operator.

By using equations (\ref{eq_temp_hetero}), (\ref{eq_vor_hetero}) and the time derivative of (\ref{eq_velocity_hetero})  with  the identity $D_xD_x u_x^{\prime} = - {D_x} {D_z} u_z^{\prime}$, the $mIG$ wave equation in axial velocity ($u_{z}^{\prime}$) is obtained as,
\begin{equation} \label{eq_operator_hetero}
\left[ \frac{\partial^2}{\partial t^2} \nabla^2 + {\left (\frac{Pr}{E}\right)}^2(\hat{\textbf{z}} \cdot \nabla)^2 + RaPr \Gamma_z^* \nabla_h^2   - RaPr \Gamma_x^* D_x D_z  \right] u_z^{\prime} = 0
\end{equation}
where, $D_x = \frac{\partial}{\partial x}, D_z = \frac{\partial}{\partial z}$.

Assuming normal mode expansion [Eqs. (\ref{eq_bi_global_perturbation}] with $y$-invariance and wave-like form for the axial velocity perturbation ($u_z^{\prime}$) given by,
\begin{equation}
\label{eq_normal_mode}
{u_z}^{\prime}(x,y,z,t) = {u_z}^{\prime}(x) e^{i(mz - \zeta t)}
\end{equation} 
in the equation (\ref{eq_operator_hetero}), 
the reduced operational form is obtained as,
\begin{equation} \label{eq_dispersion_hetero_partial}
\zeta^2 \left ( - \frac{d^2 u_z^{\prime}(x)}{d x^2} + m^2 u_z^{\prime}(x)  \right) - \left ( \frac{Pr}{E} \right)^2 m^2 u_z^{\prime}(x)  + RaPr \Gamma_z^* \frac{d^2 u_z^{\prime}(x)}{d x^2} - i m  RaPr {\Gamma_x^*} \frac{d u_z^{\prime}(x)}{d x} = 0
\end{equation}

In order to obtain a constant coefficient algebraic form for the dispersion relation with spatially heterogeneous buoyancy, the axial and lateral temperature gradients are assumed to be constant. This allows a wave-like form for the axial velocity perturbations in the $x$ direction given by $u_z^{\prime}(x) = u_0 e^{ikx}$. 

Hence, the dispersion relation is obtained as,
\begin{equation} \label{eq_dispersion_hetero_full_1}
\zeta^2 (k^2 + m^2) = \left ( \frac{Pr}{E} \right)^2 m^2 + k^2 RaPr \Gamma_z^* - m k RaPr \Gamma_x^*
\end{equation}
%, ${K_h}^2 = {k}^2$, and $k_x = k, k_z = m$ are the horizontal and vertical wave numbers, respectively. Hence, in 
The exact closed-form solutions of the above dispersion relation ($\ref{eq_dispersion_hetero_full_1}$) are obtained as,
\begin{equation} \label{eq_dispersion_hetero_full_3}
\zeta = \pm \left[  \left ( \frac{Pr}{E} \right)^2 \frac{m^2}{k^2 + m^2} + RaPr \Gamma_z^* \frac{{k}^2}{k^2 + m^2} -  RaPr \Gamma_x^* \frac{k m}{k^2 + m^2} \right]^{1/2}.
\end{equation}

\section{Penetrative magneto-convection} 
Consider equations (\ref{eq_incom_magnetic_dimensionless_perturb}) - (\ref{eq_induction_magnetic_dimensionless_perturb}) with the non-linear and diffusivity terms neglected. Hence, the linearised perturbation equations for magnetoconvection are obtained as,
\begin{equation} \label{eq_incom_magneto}
\mathbf{\nabla} \cdot \mathbf{u}^{\prime} = 0, \hspace{10pt} \mathbf{\nabla} \cdot \mathbf{B}^{\prime} = 0
\end{equation}
\begin{equation} \label{eq_NS_magneto}
\frac{\partial \mathbf{u}^{\prime}}{\partial t} + (\mathbf{u}^* \cdot \nabla) \mathbf{u}^{\prime} + (\mathbf{u}^{\prime} \cdot \nabla) \mathbf{u}^* + \frac{Pm}{E} (\hat{\textbf{z}} \times \mathbf{u}^{\prime}) = - \nabla P^{\prime} + \frac{\Lambda Pm}{E} (\mathbf{B}^* \cdot \nabla) \mathbf{B}^{\prime} + q Ra Pm T^{\prime}  \hat{\textbf{z}} ,
\end{equation}
\begin{equation} \label{eq_temp_magneto}
\frac{\partial T^{\prime}}{\partial t} + u_x^{\prime} \Gamma_x^{*} + u_z^{\prime} \Gamma_z^{*} = 0
\end{equation}
\begin{equation} \label{eq_induction_magneto}
\frac{\partial \mathbf{B}^{\prime}}{\partial t} = (\mathbf{B}^* \cdot \nabla) \mathbf{u}^{\prime}
\end{equation}
By operating ($\nabla \times $) on equation (\ref{eq_NS_magneto}) and (\ref{eq_induction_magneto}) leads to, 
\begin{equation} \label{eq_NS_magneto_curl1}
\frac{\partial \bm{\omega}^{\prime}}{\partial t} = \frac{Pm}{E} (\hat{\textbf{z}} \cdot \nabla) \mathbf{u^{\prime}}  + \frac{\Lambda Pm}{E} (\mathbf{B}^* \cdot \nabla) (\nabla \times \mathbf{B}^{\prime}) + qRaPm(\nabla \times T^{\prime} \hat{\mathbf{z}} ),
\end{equation}
\begin{equation} \label{eq_induction1_hetero_magneto}
\frac{\partial}{\partial t} (\nabla \times \mathbf{B^{\prime}}) = (\mathbf{B^*} \cdot \nabla) \bm{\omega}^{\prime}
\end{equation}
where, $\bm{\omega}^{\prime} = (\nabla \times \mathbf{u}^{\prime})$ is the vorticity. Further taking $\frac{\partial}{\partial t} (\nabla \times)$ of equation (\ref{eq_NS_magneto_curl1}) yields,
\begin{equation} \label{eq_NS_magneto_curl2}
\left[\frac{\partial^2}{\partial t^2} - \frac{\Lambda Pm}{E} (\mathbf{B^* \cdot \nabla})^2\right] \nabla^2 \mathbf{u}^{\prime} = - \frac{Pm}{E} (\hat{\textbf{z}} \cdot \nabla) \frac{\partial \mathbf{\omega}^{\prime}}{\partial t} + q Ra Pm \nabla_h^2 \left(\frac{\partial T^{\prime}}{\partial t} \right) \hat{\mathbf{z}}
\end{equation}
Implementing the identity, $\nabla_h^2 u_x^{\prime} = - {D_x} {D_z} u_z^{\prime}$, and further taking ($\hat{\mathbf{z}} \cdot \frac{\partial }{\partial t}$) of (\ref{eq_NS_magneto_curl2}) and using (\ref{eq_temp_magneto}), the wave equation in axial ($\hat{z}$) direction is obtained as,
\begin{equation} \label{eq_full_operator}
\begin{split}
\left[\frac{\partial^2}{\partial t^2} - \frac{\Lambda Pm}{E} (\mathbf{B^* \cdot \nabla})^2\right]^2 \nabla^2 u_z^{\prime} + \left[\frac{\partial^2}{\partial t^2} - \frac{\Lambda Pm}{E} (\mathbf{B^* \cdot \nabla})^2\right] \left( q Ra Pm \Gamma_z^* \nabla_h^2 - q  RaPm \Gamma_x^* D_x D_z \right) u_z^{\prime} \\ +  \left(\frac{Pm}{E}\right)^2 (\hat{\textbf{z}} \cdot \nabla)^2 \frac{\partial^2 u_z^{\prime}}{\partial t^2} = 0
\end{split}
\end{equation}
and simplifies to,
\begin{equation} \label{eq_full_operator_final}
\begin{split}
\left[\frac{\partial^2}{\partial t^2} - \frac{\Lambda Pm}{E} (\mathbf{B^* \cdot \nabla})^2\right] \left[ \left(\frac{\partial^2}{\partial t^2} - \frac{\Lambda Pm}{E} (\mathbf{B^* \cdot \nabla})^2\right)\nabla^2 +  q Ra Pm \Gamma_z^* \nabla_h^2 - q  RaPm \Gamma_x^* D_x D_z \right]u_z^{\prime} \\ +  \left(\frac{Pm}{E}\right)^2 (\hat{\textbf{z}} \cdot \nabla)^2 \frac{\partial^2 u_z^{\prime}}{\partial t^2} = 0
\end{split}
\end{equation}
The above equation represents the modified MAC wave which is substantially altered in the presence of a laterally heterogeneous temperature gradient from its homogeneous counterpart described in previous studies \cite{takehiro2015penetration, takehiro2018penetration, xu2024penetrative}. 

With the dominance of rotational effects over magnetic field effects, the approximation  $\Lambda << 1$ leads to the neglecting of the operatorial terms containing $\frac{\Lambda Pm}{E} (\mathbf{B^* \cdot \nabla})$. This leads to the rotationally influenced modified MAC wave equation (\ref{eq_full_operator_final}) to,
\begin{equation} \label{eq_full_operator_final_rapidrotation}
\left[\frac{\partial^2}{\partial t^2} \nabla^2 + \left(\frac{Pm}{E}\right)^2 (\hat{\textbf{z}} \cdot \nabla)^2 +  q Ra Pm \Gamma_z^* \nabla_h^2 - q  RaPm \Gamma_x^* \nabla_x \nabla_z \right]u_z^{\prime} = 0
\end{equation}
which has the same form as the case of mIG wave in equation (\ref{eq_operator_hetero}).

As $Pr = Pm$ implying $q = 1$ is used for all cases in the present study, the above equation becomes mathematically identical to equation (\ref{eq_operator_hetero}). 

The similarity in the mathematical form of the modified MAC wave equation (\ref{eq_full_operator_final_rapidrotation}) to that for the mIG wave [Eqs.\ref{eq_operator_hetero}] allows the use of a similar procedure [see Eqs.(\ref{eq_operator_hetero}) - (\ref{eq_dispersion_hetero_full_3})] to derive the dispersion relation of magnetically modified inertial-gravity waves (mmIG) as
\begin{equation} \label{eq_gravity_wave_dispersion_hetero_full_rot}
\zeta^2 (k^2 + m^2) = \left ( \frac{Pm}{E} \right)^2 m^2 + k^2 qRaPm \Gamma_z^* - m k qRaPm \Gamma_x^*
\end{equation}
The exact closed form solutions of the above dispersion relation ($\ref{eq_gravity_wave_dispersion_hetero_full_rot}$) is obtained as,
\begin{equation} \label{eq_dispersion_hetero_full_3a}
\zeta = \pm \left[  \left ( \frac{Pm}{E} \right)^2 \frac{m^2}{k^2 + m^2} + qRaPm \Gamma_z^* \frac{{k}^2}{k^2 + m^2} -  qRaPm \Gamma_x^* \frac{k m}{k^2 + m^2} \right]^{1/2}.
\end{equation}

With strong magnetic field dominating over rotational effects, the assumption $\Lambda >> 1$ in equation (\ref{eq_full_operator_final}) results in
\begin{equation} \label{eq_full_operator_magnetic}
\underbrace{\left[\frac{\partial^2}{\partial t^2} - \frac{\Lambda Pm}{E} (\mathbf{B^* \cdot \nabla})^2\right]}_{\text{First Operator}} \underbrace{\left[ \left(\frac{\partial^2}{\partial t^2} - \frac{\Lambda Pm}{E} (\mathbf{B^* \cdot \nabla})^2\right)\nabla^2 +  q Ra Pm \Gamma_z^* \nabla_h^2 - q  RaPm \Gamma_x^* D_x D_z \right]}_{\text{Second Operator}} u_z^{\prime}  = 0
\end{equation}
The first operator in the above equation represents the pure Alfv\'en wave, which does not depend on the stratification\cite{takehiro2015penetration,xu2024penetrative}. However, the second operator represents the magnetically modified gravity (mmG) wave equation which depends on the strength of stratification.

Therefore, the mmG wave equation is obtained by disregarding the solutions obtained from the first operator from (\ref{eq_full_operator_magnetic}). This is essentially equivalent to retaining only the second operator to get,
\begin{equation} \label{eq_full_operator_MAC}
\left[ \left(\frac{\partial^2}{\partial t^2} - \frac{\Lambda Pm}{E} (\mathbf{B^* \cdot \nabla})^2\right)\nabla^2 +  q Ra Pm \Gamma_z^* \nabla_h^2 - q  RaPm \Gamma_x^* D_x D_z \right]u_z^{\prime}  = 0
\end{equation}
By using equation (\ref{eq_normal_mode}) and implementing a wave-like form for the axial velocity perturbations in the $x$ direction given by $u_z^{\prime}(x) = u_0 e^{ikx}$ on equation (\ref{eq_full_operator_MAC}), the closed form solutions to the resulting dispersion relation is obtained as,
\begin{equation} \label{eq_full_operator_MAC_dispersion}
\zeta = \pm \left[  \left ( \frac{\Lambda Pm}{E} \right)^2 \left(\mathbf{B}^* \cdot \mathbf{K} \right)^2 + qRaPm \Gamma_z^* \frac{{k}^2}{k^2 + m^2} -  qRaPm \Gamma_x^* \frac{km}{k^2 + m^2} \right]^{1/2}
\end{equation}
By setting buoyancy contributions, $\Gamma_z^* = \Gamma_x^* = 0$, the dispersion relation reduces to pure Alfv\'en wave frequency as $\zeta_A = \pm \left[ \left(\frac{\Lambda Pm}{E} \right) (\mathbf{B}^* \cdot \mathbf{K}) \right]$. For waves propagating perpendicular to the imposed magnetic field ($\mathbf{B}^* \cdot \mathbf{K} = 0$), retaining the buoyancy forcing leads to $mmG$ waves. Since the wave frequency (Eq. \ref{eq_full_operator_MAC_dispersion}) depends on the wave number and direction of the imposed magnetic field, therefore, these waves are dispersive and anisotropic. Furthermore, fast $mmG$ wave frequency is obtained for $\zeta > \zeta_A$ and slow $mmG$ wave occurs when $\zeta < \zeta_A$. 
\bibliographystyle{acm}
\bibliography{reference}
\end{document}